%% file: main.tex
\documentclass[twocolumn]{aastex63}
\usepackage{xcolor}
\usepackage{amsmath}
\usepackage{rotating}
\usepackage[utf8]{inputenc}
\usepackage{array}
\usepackage{makecell,gensymb}

\newcommand{\sn}{SN\,2004C}
\newcommand{\dist}{35.9\,Mpc}
\newcommand{\chev}{\citet{chev98}}
\newcommand{\kms}{\mbox{km s$^{-1}$}}
\newcommand{\sm}{\mbox{$M_{\sun}$}}
\newcommand{\WRwind}{1000\,\kms}
\newcommand{\constML}{$10^{-5}$\,\sm\,yr$^{-1}$}
\newcommand{\seti}{ \overset{\tiny i=5/3}{=}}
\newcommand{\setq}{ \overset{\tiny q=7/8}{=}}
\newcommand{\setg}{ \overset{\tiny \Gamma=5/3}{=}}
\newcommand{\host}{NGC\,3683}
\newcommand{\vlbaMeas}{$3.8^{+0.6}_{-1.0}\times 10^{17}$}

\newcommand{\fitcase}{$s = -1, \alpha_2=5/2$}  
  
\newcommand{\thin}{$-0.96 \pm 0.03$}  
  
\newcommand{\pvalue}{$2.92 \pm 0.06$}  
\newcommand{\firstrhoslope}{$-0.03 \pm 0.22$}  
\newcommand{\secondrhoslope}{$-2.3 \pm 0.5$}  

\newcommand{\q}{$0.77 \pm 0.04$}  
\newcommand{\qVLBA}{$0.81 \pm 0.03$} 
\newcommand{\firstbvsrslope}{$-0.13 \pm 0.08$}  
\newcommand{\secondbvsrslope}{$-1.10 \pm 0.07$}

\newcommand{\lowwind}{2\,\kms}  
  
\newcommand{\lastwind}{18\,\kms}

\newcommand{\maxmassloss}{$(460 \pm 80) \times 10^{-5}$\,\sm\,yr$^{-1}$}

\newcommand{\MLinner}{$(54 \pm 8)\times10^{-5}$\,\sm\,yr$^{-1}$}

\newcommand{\lowestUequi}{$(1.60\pm0.25)\times10^{47}$}

\newcommand{\highestUequi}{$(1.9\pm0.4)10^{48}$}
\newcommand{\dayshighestUequi}{242 days}

\newcommand{\innerradius}{$(8.1 \pm 0.5)\times 10^{15}$}  
\newcommand{\outerradius}{$(3.4 \pm 0.4)\times 10^{16}$}  
\newcommand{\breakradius}{$(1.96 \pm 0.10)\times 10^{16}$}  
  
\newcommand{\VLBApredR}{$(1.5 \pm 0.5)\times 10^{17}$}

\newcommand{\timeWRwindInner}{$10^{2.97}$ days (2.6\,yr)}  
\newcommand{\timeWRwindOuter}{$10^{3.59}$ days (10.8\,yr)}  
\newcommand{\timebreakMLRadius}{$10^{3.36}$ days (6.2\,yr)}

\newcommand{\shellsize}{0.021\,\sm}  
\newcommand{\percentlife}{$10^{-5} - 0.01\%$}

\newcommand{\berkeley}{Department of Astronomy and Astrophysics, University of California, Berkeley, CA 94720, USA}
\newcommand{\ciera}{Center for Interdisciplinary Exploration and Research in Astrophysics and Department of Physics and Astronomy, Northwestern University, 2145 Sheridan Road, Evanston, IL 60208-3112, USA}
\newcommand{\york}{Department of Physics and Astronomy, York University, Toronto, M3J 1P3, Ontario, Canada}
\newcommand{\purdue}{Department of Physics and Astronomy, Purdue University, 525 Northwestern Avenue, West Lafayette, IN 47907, USA}
\newcommand{\idsi}{Integrative Data Science Initiative, Purdue University, West Lafayette, IN 47907, USA}
\newcommand{\mpifr}{Max-Planck-Institut f\"ur Radioastronomie, Auf dem H\"ugel 69, 53121 Bonn, Germany}
\newcommand{\mpheidel}{Max Planck Institute for Astronomy, Konigstuhl 17, D-69117 Heidelberg, Germany}
\newcommand{\warwick}{Department of Physics, University of Warwick, Gibbet Hill Road, Coventry CV4 7AL, UK}
\newcommand{\ucsc}{Department of Astronomy and Astrophysics, University of California, Santa Cruz, CA 95064, USA}
\newcommand{\umel}{School of Physics, The University of Melbourne, VIC 3010, Australia}
\newcommand{\astrothreeD}{ARC Centre of Excellence for All Sky Astrophysics in 3 Dimensions (ASTRO 3D)}
\newcommand{\cfa}{Center for Astrophysics, Harvard \& Smithsonian, 60 Garden Street, Cambridge, MA 02138, USA}

\usepackage{graphicx}
\usepackage[caption=false]{subfig}
\usepackage{breqn}

\received{\today}

\submitjournal{ApJ}

\shorttitle{Radio Analysis of \sn}
\shortauthors{DeMarchi et al.}

\graphicspath{{./}{figures/}}

\begin{document}

\title{Radio Analysis of \sn\, Reveals an Unusual CSM Density Profile as a Harbinger of Core Collapse}

\author[0000-0003-4587-2366]{Lindsay ~DeMarchi} \affil{\ciera}
\author[0000-0003-4768-7586]{R. Margutti} \affil{\berkeley}
\author{J. Dittman} \affil{\cfa} \affil{\mpheidel}
\author[0000-0003-4468-761X]{A. Brunthaler}\affil{\mpifr}
\author[0000-0002-0763-3885]{D. Milisavljevic} \affil{\purdue}
\affil{\idsi}
\author[0000-0002-0592-4152]{Michael F. Bietenholz} \affil{\york}
\author[0000-0001-8769-4591]{C. Stauffer} \affil{\ciera}
\author[0000-0001-6415-0903]{D. Brethauer} \affil{\berkeley}
\author[0000-0001-5126-6237]{D. Coppejans} \affil{\warwick}\affil{\ciera}
\author[0000-0002-4449-9152]{K.~Auchettl} \affil{\ucsc} \affil{\umel} \affil{\astrothreeD}
\author[0000-0002-8297-2473]{K. D. Alexander} \affil{\ciera}
\author[0000-0002-5740-7747]{C. D. Kilpatrick}\affil{\ciera}
\author[0000-0002-7735-5796]{Joe S. Bright}\affil{\berkeley}
\author[0000-0002-6625-6450]{L. Z. Kelley} \affil{\ciera}
\author[0000-0002-3019-4577]{Michael C. Stroh} \affil{\ciera}
\author[0000-0002-3934-2644]{W.~V.~Jacobson-Gal\'an} \affil{\berkeley}


\begin{abstract}
We present extensive multi-frequency VLA and VLBA  observations of the radio-bright supernova (SN) IIb \sn\ that span ${\sim}$40--2793 days post-explosion. We interpret the temporal evolution of the radio spectral energy distribution (SED) in the context of synchrotron self-absorbed (SSA) emission from the explosion's forward shock as it expands in the circumstellar medium (CSM) previously sculpted by the mass-loss history of the stellar progenitor. VLBA observations and modeling of the VLA data point to a blastwave with average velocity $\sim0.06 c$ that carries an energy of ${\approx}10^{49}$ erg. 
Our modeling further reveals a flat CSM density profile $\rho_{\rm CSM}{\propto}R^{\text{\firstrhoslope}}$  up to a break radius $R_{\rm br}\approx $\breakradius\ cm, with a steep density gradient following $\rho_{\rm CSM}{\propto}R^{\text{\secondrhoslope}}$  at larger radii. 
We infer that the flat part of the density profile corresponds to a CSM shell with mass $\sim$\shellsize, and that the progenitor's effective mass-loss rate varied with time over the  range $(50 - 500) \times 10^{-5}$ \sm\ yr$^{-1}$
for an adopted wind velocity $v_w=$1000 \kms\ and shock microphysical parameters $\epsilon_e=0.1$, $\epsilon_B=0.01$. These results add to the mounting observational evidence for departures from the traditional single-wind mass-loss scenarios in evolved, massive stars in the centuries leading up to core collapse. 
Potentially viable scenarios include mass loss powered by gravity waves and/or interaction with a binary companion. 
\end{abstract}

\section{Introduction} 
\label{sec:intro}

All stars lose mass to their environments throughout their lifetimes, critically affecting both the star's evolution and fate in addition to enriching the Universe with heavy elements. If the star dies in a supernova (SN) explosion, shock interaction with recent mass lost to the circumstellar medium shapes the appearance of the SN display. 

In recent years, observations of supernovae (SNe) have uncovered evidence that evolved massive stars live complex lives with a rich and diverse mass-loss history, particularly in the final moments before their deaths.
Large optical datasets that sampled the pre-explosion phase of SNe (such as the targets observed by the Palomar Transient Factory -PTF- and the Zwicky Transient Facility -ZTF-; \citealt{law09, bellm19, graham19}) have unveiled evidence of pre-explosion eruptions in $>50\%$ of progenitors of Type-IIn SNe (\citealt{ofek14}, later updated by \citealt{strot21}),  which are SNe that explode into dense 
H-rich CSM and show narrow hydrogen lines in their spectra as a consequence (\citealt{filippenko97}).
Among Type-IIn SNe, the precursor event of SN\,2010mc \citep{ofek13} and the long history of pre-SN eruptions of SN\,2009ip (\citealt{Smith1009ip, Foley11}) were among the first events inferring the presence of a dense CSM shell extending $\sim5\times10^{14} - $ $\sim4\times10^{16}$ cm from their explosion sites. 
The existence of these shells points to a mass-loss mechanism different from the traditional line-driven winds, and hence exposes gaps in our understanding of massive stellar evolution \citep{Fraser13, Mauerhan13a, Mauerhan14, Pastorello13, Prieto13, margutti09ip, smitharnett14}.  
Interestingly, evidence of large mass-loss events ($0.01 -$ several $\sm$ lost to the environment) occurring in the centuries to years before death 
extends to the H-rich progenitors of superluminous SNe (SLSNe-II), a class assigned to SNe orders of magnitude more luminous than normal ones that also show H in early spectra. SN\,2008es is one such example, which interacted with $2-3M_{\sun}$ material 0.5-1.6 years prior to explosion (\citealt{bhiro19}).

Importantly, violent pre-SN mass loss is not strictly limited to  H-rich progenitors. Revealed by an increasing wealth of multi-wavelength observational evidence, H-\textit{poor} progenitors of SNe Type IIb and Type Ib/c (SN Ib/c) exhibit enhanced mass loss prior to explosion as well 
(see e.g., \citealt{mauerhan18, kundu19, prentice20, gomez19, jin21, guti21, maeda21, benetti18, kuncara18, wynn19ehkearly, wynn19ehk}, for recent examples).
This phenomenon also extends to H-poor SLSNe (SLSNe-I). For example, iPTF13ehe, iPTFesb, iPTF16bad, iPTF16eh, SN\,2018bsz, and other events discussed in \citet{Griffin2021} show evidence of CSM material located $\sim 10^{16}$cm from their explosion sites (\citealt{yan17, lunnan18, anderson18, chen21, pursiainen22}).

Radio observations of SNe played a key role in our understanding of ``unusual'' CSM density profiles around SNe (e.g., \citealt{chevfrans17} for a recent review). In a SN explosion, while optical observations sample the slowly expanding ejecta 
($v \leq 10^4$ \kms)\ radio observations probe the fastest ejecta ($v \geq 0.1 c$). 
Radio synchrotron emission originates from the interaction of the fastest SN ejecta with the local CSM deposited by the progenitor star prior to explosion. 
By monitoring the temporal evolution of the spectral peak frequency and flux density, one may directly constrain physical properties of the system, such as the mass-loss rate $\dot M$ of the progenitor star. 
Therefore, radio observations allow us to map the density profile of the CSM around the explosion site.
Previous works in the literature have built a foundation of using this approach to measure pre-explosion mass loss of H-stripped progenitors at radio frequencies as the blastwave interacts with the surrounding CSM (see e.g., SNe\, 2001em, 2003bg, 2003gk, 2007bg, 2008ax, and PTF11qcj; \citealt{soderberg03bg, bieten14, salas13, schinzel09, roming09, Corsi14}).
Using the standard techniques of radio analysis and radio data modeling (as reviewed in \citealt{weiler02, chevfrans17}), we have found that \sn\, is one of such H-poor explosions with a shockwave that encounters a dense CSM with a profile that is not consistent with the traditional single stellar wind models.  

Observations across the electromagnetic spectrum of SNe  thus point to the presence of strongly time-dependent mass loss predicating core collapse in a diversity of explosion types, ranging from stripped-envelope SNe Ib/c to SNe II and extending to SLSNe.
The fact that these mass-loss ejections occur across different types of SNe implies an origin of the outbursts that is independent of progenitor size.
The exact nature of the underlying physical process causing this time-dependent mass loss remains the subject of debate, as well as how these processes inform the evolutionary path of the progenitor (e.g., \citealt{smith2014}).

In this work we analyze the radio data of \sn, a SN Type IIb that exploded December 15, 2003 (MJD 52988, see \S\ref{sec:optical}) located in the galaxy NGC 3683, \dist\ away \citep{Dudley04C, Tully2009, tully13}. 
Our radio dataset spans $40-2793$ days post explosion and represents one of the most extensive radio datasets of a SN thus far, which we model with synchrotron self-absorption. Furthermore, Very Long Baseline Array (VLBA) imaging of \sn\ at 1931 days post explosion offers a direct measurement of the blastwave radius, 
making \sn\ among the $\sim10$ SN with published VLBA constraints to date. 
With the use of our generalized equations from \chev\ (hereafter C98), we find the explosion site of \sn\ contains a dense shell of CSM material and we speculate on its physical origin.

Our paper is organized as follows.
The VLA and VLBA data of \sn\ are presented in \S\ref{sec:obs}, and our generalization of the equations following the formalism of C98 is in \S\ref{sec:derivation}. Our modeling of the data and analysis of the physical parameters of the system are presented in  \S\ref{sec:phys}.
Our comparison to plausible mass-loss mechanisms are presented in  \S\ref{sec:massloss}, and we summarize our results in \S\ref{sec:conclusion}.

\section{Observations} \label{sec:obs}

\input{sn2004c_v2}

\subsection{Karl G. Jansky Very Large Array}

We observed \sn\ with the NSF's Karl G. Jansky Very Large Array (VLA) beginning on January 23.65, 2004 UT, $\delta t \approx 30$ days after the initial discovery under programs AK0575 and AS0796 (PI Soderberg). We found a radio source at $\alpha = 11^h27^m29.72^s$, $\delta = +56{\degree}52'48.2''$, coincident with the optical SN position with a flux density of $F_{\nu} = 1.6 \pm 0.05$ mJy at  a frequency of $\nu=8.5$ GHz. We continued to monitor this source for the next $\delta t \approx 2800$ days.

Radio data were taken in the 4.9, 8.5, 15.0, and 22.0 GHz bands, and we give the measured flux densities in Table \ref{tab:data}. Observations were taken in continuum observing mode with $2\times50$ MHz bandwidth, except that in August 2011, which was  taken with the upgraded EVLA system, with an increased 2 GHz bandwidth. We used observations of J1127+568 to calibrate the time dependent complex gains of the instrument and 3C286 to calibrate the absolute flux scale and the bandpass response for all observations. Data were reduced using the Astronomical Image Processing System (AIPS) by fitting a elliptical Gaussian model to the radio SN in each observation to measure the integrated flux density. We note that that \sn\ has a peak luminosity density at 100 days of $\sim 10^{28}$ erg~s$^{-1}$ Hz$^{-1}$ and is at the high end of the radio luminosity function for SNe \citep[see, e.g.,][]{Bietenholz+2021}. A subsection of data points showcasing the SN evolution is shown in Figure \ref{fig:SED_evol} and the full dataset is plotted in Appendix Figure \ref{fig:SED}. In Figure \ref{fig:freqslices}, we show the lightcurves for \sn\ at 4.9, 8.5, 15.0, and 22.0 GHz bands.

\begin{figure*}[!ht]
    \centering
    \includegraphics[width=100mm]{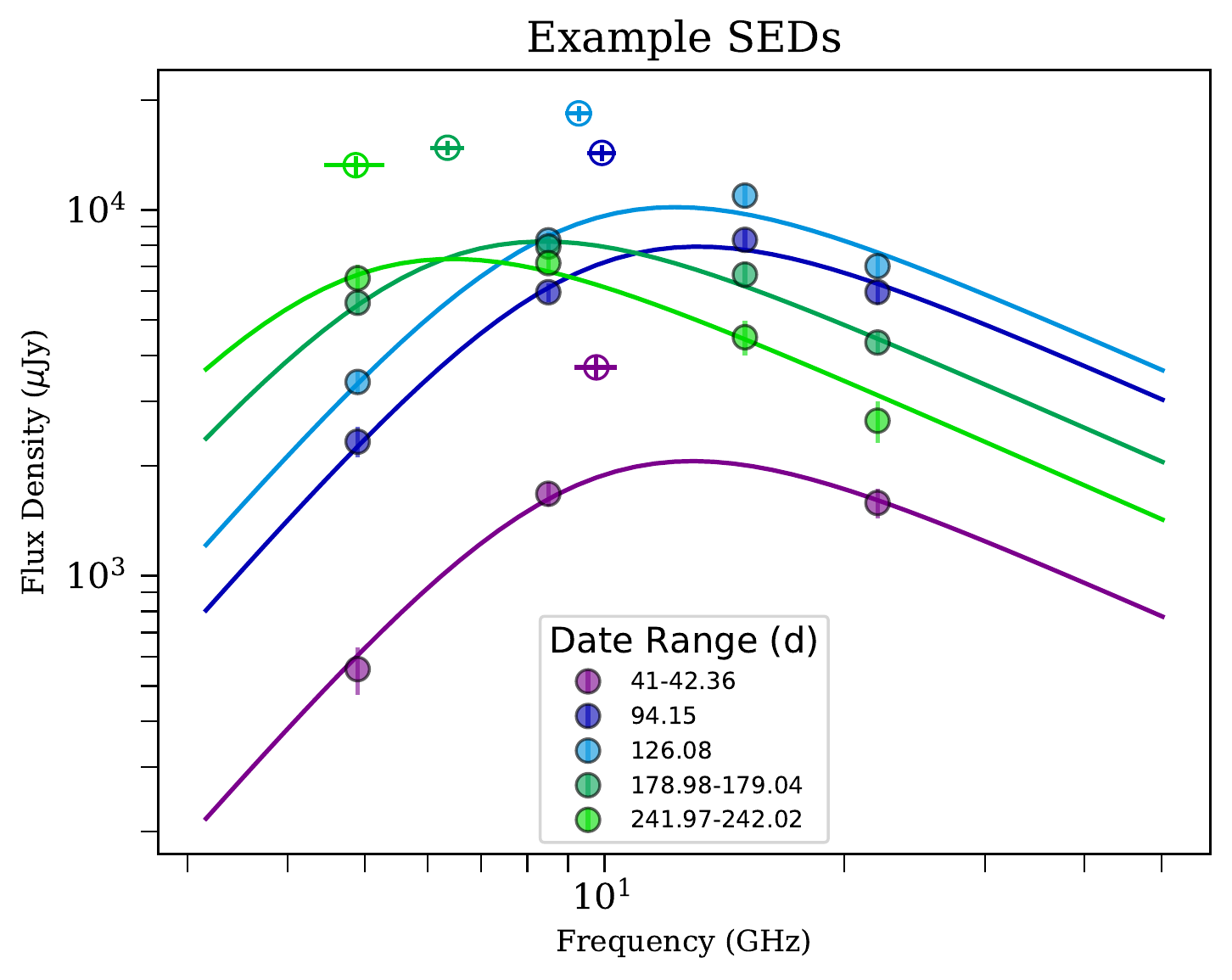}
    \caption{A few example radio SEDs are chosen to show the evolution of the emission through time. 
    Note that for \sn\, the evolution of the peak flux density and peak frequency does not follow a power law with time
    and our modeling is designed to adapt to this non-standard evolution.
   \textbf{ Solid lines:} best-fitting model. \textbf{Solid points:} radio observations of \sn. \textbf{Hollow points:} intercepts of the asymptotic power-law segments, which mark the location of $\nu_{\rm{brk}}$ and  $F_{\rm{brk}}$. These values are used to estimate the physical properties of the SN outflow and its environment. }
    \label{fig:SED_evol}
\end{figure*}

\begin{figure*}[!ht]
\begin{tabular}{cc}
    \includegraphics[width=65mm]{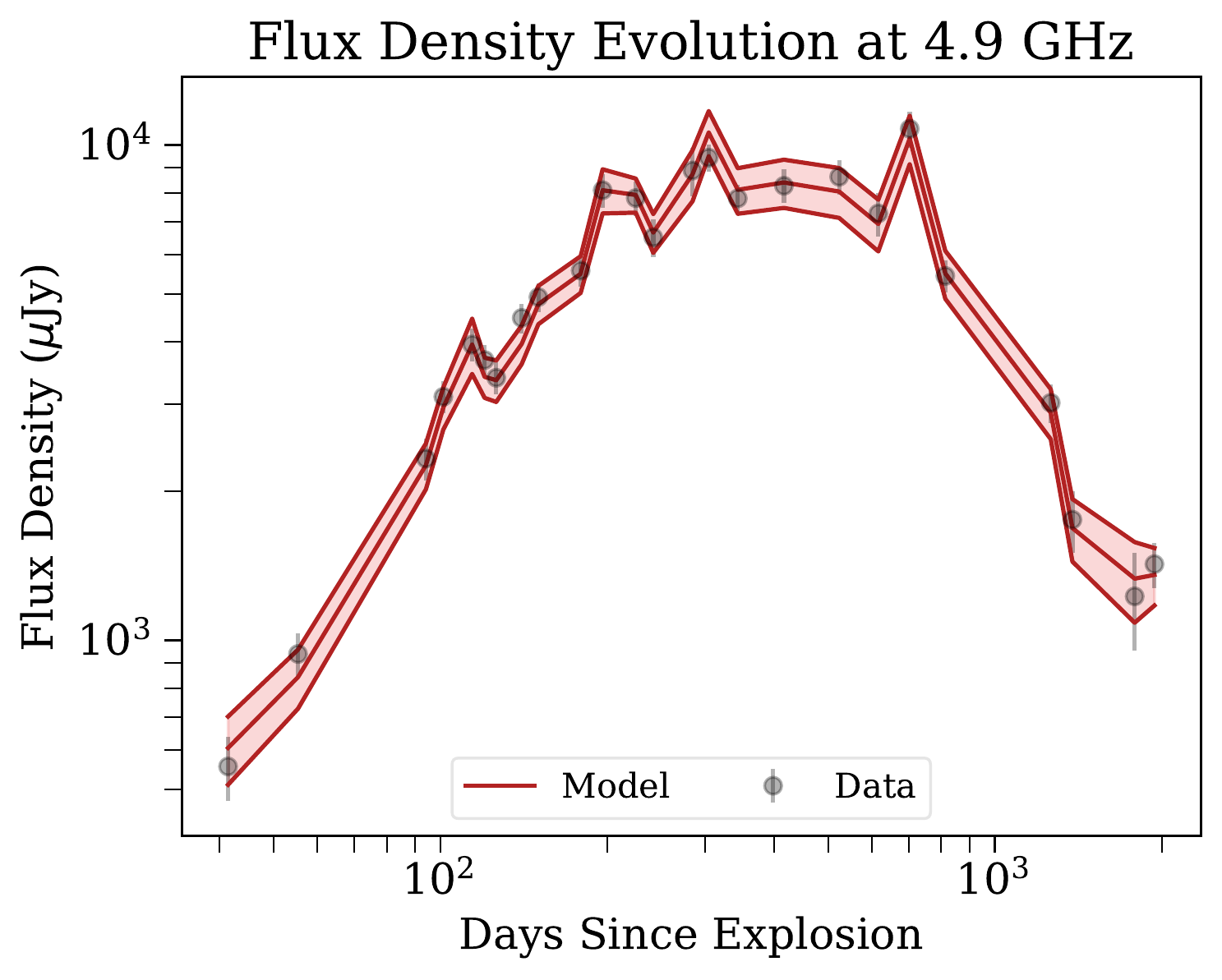} &
   \includegraphics[width=65mm]{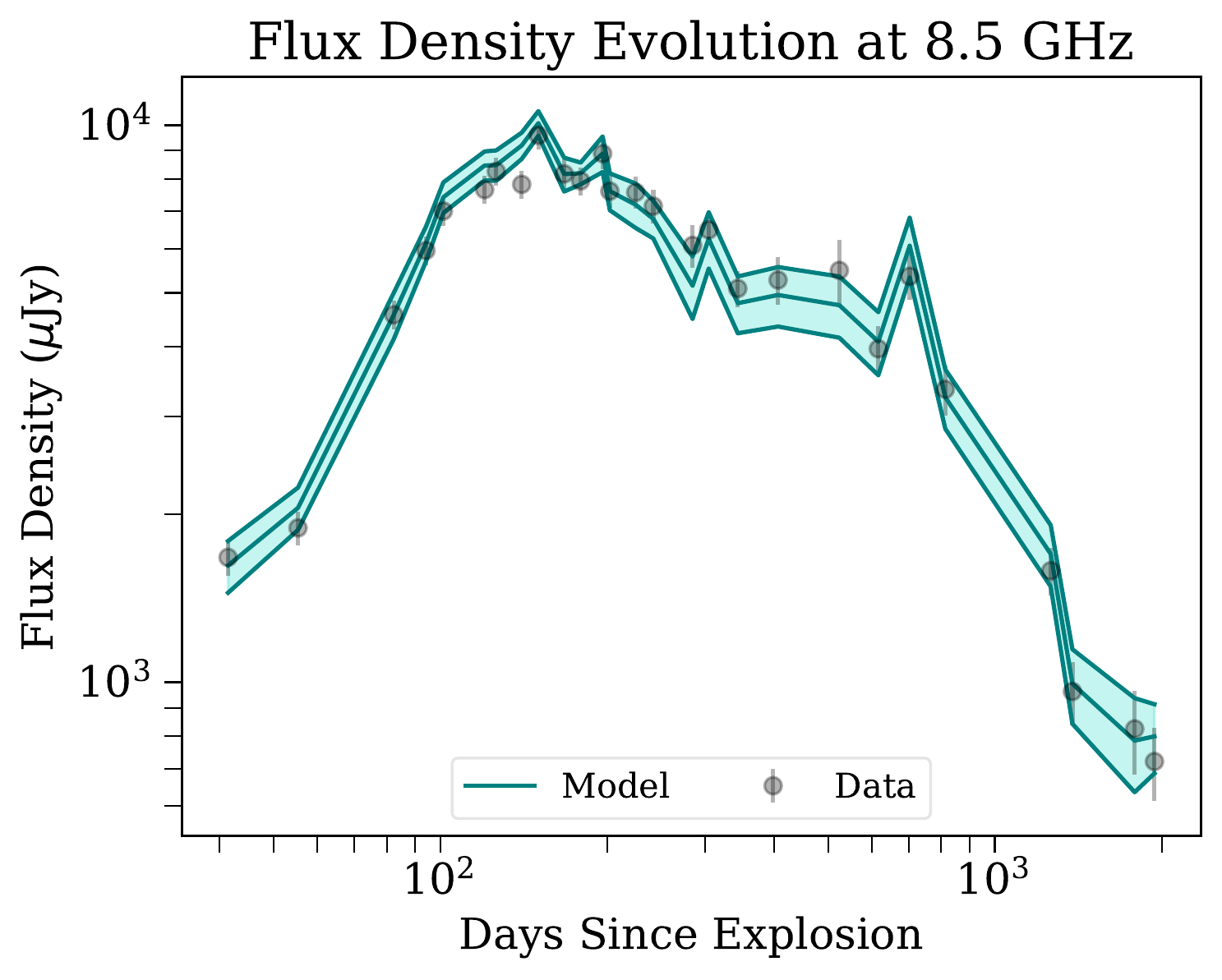}  
     \\
     \includegraphics[width=69mm]{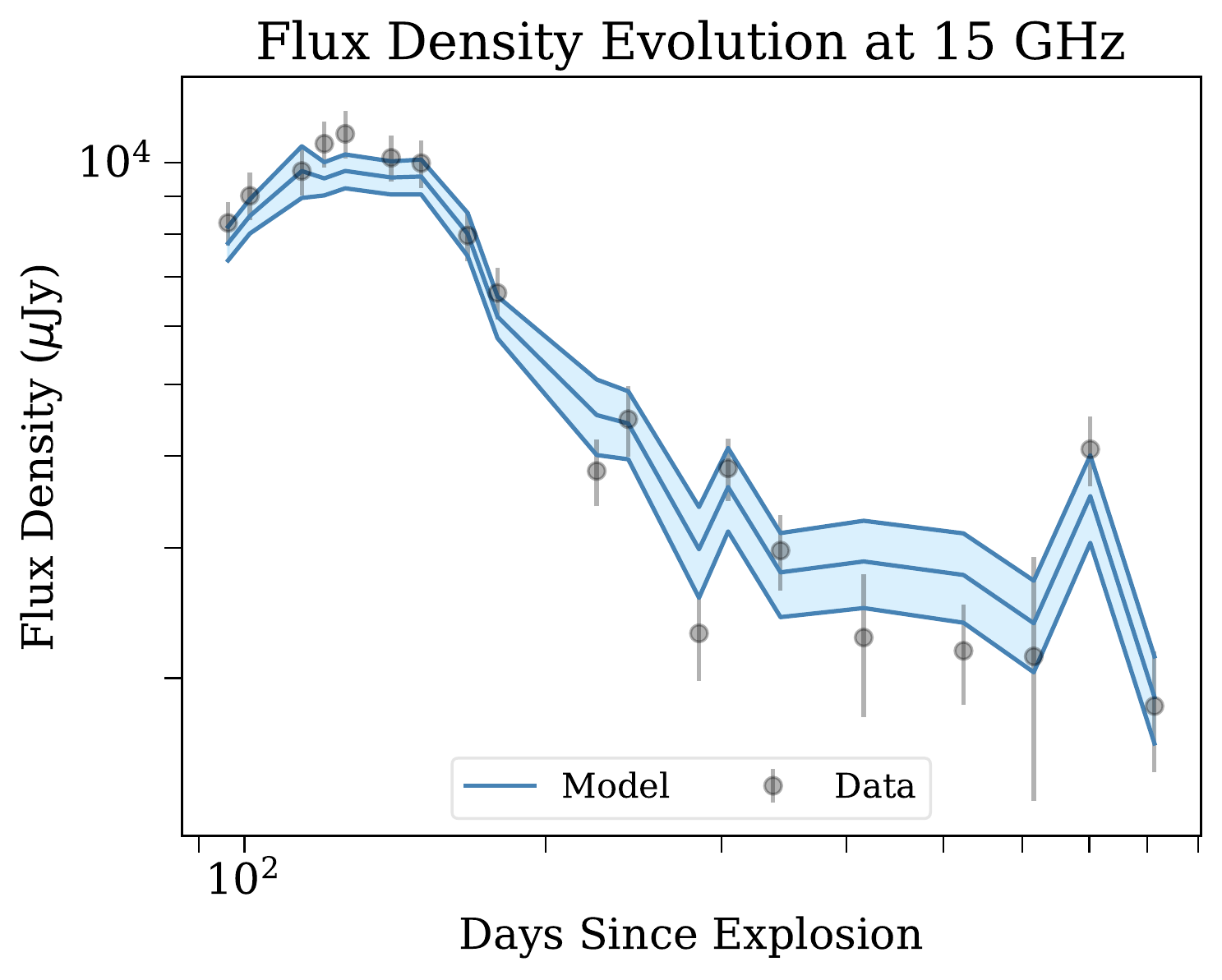}  &
  \includegraphics[width=65mm]{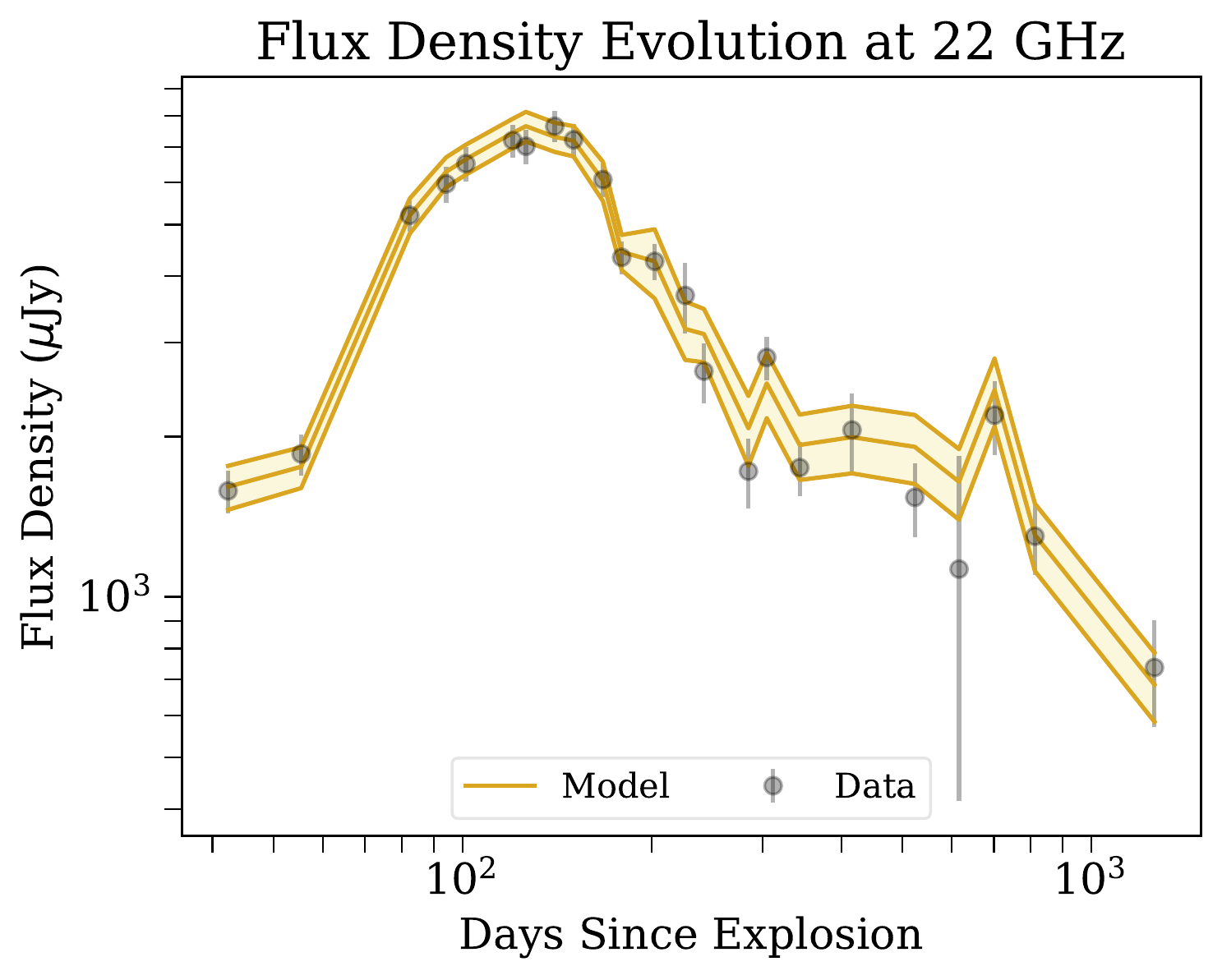} \\
\end{tabular}
\caption{Light curves of  \sn\ data at 4.9, 8.5, 15, and 22 GHz (grey points). Overlaid is our model fit to the data; shaded regions represent the spread of the $1\sigma$ uncertainty on the broken power law fit parameters (Table \ref{tab:fit}). 
}
\label{fig:freqslices}
\end{figure*}

Temporally, the radio flux density at 8.5, 15.0, and 22.0 GHz rises slowly over the first 100 days and then steadily declines. At 4.9 GHz, the radio flux density increases for approximately the first 200 days, peaking at 10 mJy, and then plateaus for several hundred more days until beginning to decline. We see evidence for a low frequency turnover in our spectral energy distribution, which is readily apparent in Figure \ref{fig:SED} of the Appendix, as well as in the raw radio lightcurves in Figure \ref{fig:freqslices}.

\subsection{Very Long Baseline Array}

We observed \sn\ with the NRAO Very Long Baseline Array (VLBA) as part of program BS192 (PI Soderberg) on March 29, 2009 at 8.4 GHz. The observations were performed with eight frequency bands of 8 MHz bandwidth each in dual circular polarization, resulting in a total data rate of 512 Mbps.  ICRF J112813.3+592514, located at $\alpha_{2000} =11^h28^m13.3406^s$ and $\delta_{2000} =+59\degree25'14.798''$ \citep{Fey04} was used for phase-referencing at both frequencies. 
We used cycle time of $\sim$4 min on \sn\ and 1 min on ICRF J112813.3+592514, and the total length of the observations was 8 hours.

The data were correlated at the VLBA Array Operations Center in Socorro, New Mexico and calibrated using AIPS and ParselTongue \citep{Kettenis06}. Total electron content maps of the ionosphere were used to correct for ionospheric phase changes. Amplitude calibration used system temperature measurements and standard gain curves. We performed a ``manual phase-calibration'' using the data from one scan of J112813.3+592514 to remove instrumental phase offsets among the frequency bands. We then fringe-fitted the data from ICRF J112813.3+592514. Finally, the calibration was transferred to \sn. The data were imaged in AIPS using robust weighting (with AIPS \texttt{ROBUST=0}).
We clearly detect the source with an integrated flux density of $F_{\nu}=0.74 \pm 0.13$ mJy at coordinates $\alpha= 11^h27^m29.728309^s \pm 0.000012^s$, $ \delta = 56\degree 52' 48.12937'' \pm 0.00014''.$ The beam size of our observation is $\approx 3$ mas, and the source is clearly resolved in our data with a size of $0.707^{+0.120}_{-0.178}$\,mas (\vlbaMeas\ cm at \dist). Our flux density of $F_{\nu} = 0.74 \pm 0.13$\,mJy is consistent with our VLA flux density measurement taken on April 5.22, 2009 UT (7 days after the VLBA measurement) of $0.72 \pm 0.07$\,mJy  (Table \ref{tab:data}).

\section{Radio Modeling}
\label{sec:derivation}

\subsection{Radio Synchrotron Emission in SNe: emitting radius $R$, post-shock magnetic field $B$, and energy $U$}
\label{SubSec:SynchRBU}

In a SN explosion, the propagation of the fastest ejecta into the environment creates a double-shock structure: one shock propagates outward (the forward shock, or FS) and the other propagates inward in mass coordinates into the SN 
ejecta (the reverse shock, or RS). The two shocked regions of CSM and SN ejecta are separated by a contact discontinuity.

Radio emission in young non-relativistic SNe is dominated by  synchrotron radiation from electrons accelerated at the FS  \citep[e.g.,][]{chev98,weiler02, chevfrans17}.
Relativistic electrons gyrate in a magnetic field $B$ with pitch angle $\theta$ in the shocked CSM and radiate photons as non-thermal synchrotron emission \citep{RyLi}. 
The acceleration of these electrons creates a power law distribution in electron Lorentz factor $\gamma_e$, where  $dN(\gamma_e)/d\gamma_e = K_0 \gamma_e^{-p}$ for $\gamma_e \ge \gamma_{\rm{m}}$,  where $\gamma_{\rm m}$ is the minimum Lorentz factor. Here,
$N$ is the number of electrons per unit volume,  $K_0$ is the normalization constant, and $p$ is the power-law index of the electron distribution.\footnote{We note that the power-law variable $p$ is sometimes expressed as $\gamma$ in the radio SN literature. 
However, we wish to avoid confusion with the electron Lorentz factor, and adopt $p$ instead.}
Therefore, the typical electron Lorentz factor of the power-law distribution is  $\gamma_e \approx \gamma_{\rm m}$.
Radio observations of SNe  typically indicate $p\approx 3$ \citep[e.g.,][for observational results, and a theoretical basis is given in \citealt{Diesing21}]{weiler02, Berger02, soderberg03bg, 
Soderberg04c,
Soderberg06c,Soderberg06d,  Soderberg07, Soderberg10a, Soderberg10b, Soderberg12,
chevandfrans,  
Corsi14, Corsi16, Kamble16}.
Diversity among the values of $p$ in SN explosions might be a manifestation of different physical properties in the shocks where particle acceleration happens.

The resulting radio flux density spectrum, $F_{\nu}$, is a series of broken power laws with spectral breaks that occur at frequencies and flux densities set by the physical properties of the system (e.g., $B$, electron density $n_e$, etc.).
Notable break frequencies are the synchrotron cooling frequency, $\nu_c$, the synchrotron characteristic frequency, $\nu_{\rm m}$, and the synchrotron self-absorption frequency, $\nu_{\rm{sa}}$. 
The slopes of $F_{\nu}$ power-law segments depend on the order of $\nu_c$, $\nu_{\rm m}$, and $\nu_{\rm{sa}}$ \citep[e.g., Fig.\ 1 of][]{granot2002}. 

We further define the shock microphysical parameters $\epsilon_B$ and $\epsilon_e$ as the fraction of the energy density $\frac{2}{1+i}\rho_{\rm{CSM}}v^2$ that goes into the post-shock magnetic field energy density $E_B$ and the relativistic electron energy density $E_e$.  We introduce the numerical factor $i$ to parametrize the different definitions of the reference energy density that appear in the literature (see Appendix \ref{App:microphysics} for details).
For our modeling, we keep the dependence on $i$ explicit. For reference, the traditional parametrization of C98 corresponds to $i=1$. 
We note that in the SN literature there are at least three other definitions of the shock microphysical parameters that we discuss in detail in Appendix \ref{App:microphysics}. It is important to keep in mind that these different definitions lead to different values of inferred parameters (most notably the mass loss, $\dot M$) even when the same shock microphysical parameters values are assumed (\S\ref{SubSec:Mdot}), as we also show in Appendix \ref{App:microphysics}.
The energy densities are defined as follows:  

\begin{equation}
\label{Eq:EBdef}
 E_B\equiv \frac{B^2}{8\pi} = \Big( \frac{2}{1+i} \Big) \epsilon_B \rho_{\rm{CSM}}v^2
\end{equation}

and

\begin{equation}
\label{Eq:Eedef}
E_e\equiv \int_{\gamma_m}^{\infty} (\gamma_e m_e c^2) K_0 \gamma_e^{-p} d\gamma_e = \Big( \frac{2}{1+i} \Big) \epsilon_e \rho_{\rm{CSM}}v^2,
\end{equation}

where $\rho_{\rm{CSM}}$ is the \emph{unshocked} CSM density and $v\equiv dR/dt$ is the FS velocity.

The frequency above which electrons cool efficiently through synchrotron radiation over a dynamical time-scale $t_{\rm dyn}$ is $\nu_c$, defined as \citep{RyLi}:
\begin{equation}
\label{Eq:nuc}
    \nu_c \equiv \frac{18 \pi m_{e} c e}{t_{\rm dyn}^2 \sigma_T^2 B^3},
\end{equation}

\noindent
where $\sigma_T$ is the Thomson cross section, $m_e$ is the mass of the electron, $e$ the charge of the electron, and $c$ is the speed of light, all in c.g.s.\ units. For our purposes $t_{\rm dyn}$ can  be identified as the time since the SN explosion. In radio SNe, $B\equiv B(t)$, which adds another dependency of  $\nu_c $ on time.

The single-electron synchrotron spectrum $F_{\nu}$ peaks at $\nu = \gamma_e^2 ( \frac{eB}{2\pi m_e c})$ \citep[e.g.,][] {RyLi},
which allows us to define the characteristic synchrotron frequency:

\begin{equation}
\label{vm}
 \nu_{\rm m}  \equiv  \gamma_m^2 \bigg( \frac{eB}{2\pi m_e c} \bigg)
\end{equation}

\noindent
Following \citet{aliciasthesis}, under the assumptions that particle kinetic energy is conserved across the shock discontinuity and a pure H composition: 

\begin{equation}
    \epsilon_e(v/c)^2m_pc^2 \approx \bar{\gamma}m_ec^2,
\end{equation}
\noindent
where $\bar{\gamma}$ is the average electron Lorentz factor and $m_p$ is the proton mass. 
For a $dN/d \gamma_e$ power-law distribution of electrons $\bar{\gamma} \approx \gamma_m$, which, substituting into Equation \ref{vm}, leads to: 
\begin{equation}
    \label{Eq:num}
    \nu_{\rm m}  \approx \epsilon_e^2 \bigg( \frac{v}{c} \bigg)^4 \bigg(\frac{m_p}{m_e} \bigg)^2 \bigg( \frac{eB}{2\pi m_e c} \bigg). 
\end{equation}
\noindent
For typical radio SN parameters \citep[e.g., from][]{soderberg03bg}, $B\sim 1$ G, $v\sim0.1\,c$,  it follows from Equations \ref{Eq:nuc} and \ref{Eq:num}  that $\nu_{c}\gtrsim 100$ GHz  and  $\nu_{\rm m}\ll1$ GHz. Taking SN\,2003bg  as an example, at $t^* \approx 35$~d since explosion \citet{soderberg03bg} infer $v(t^*)\approx0.13\,c$ 
and $B(t^*)\approx 0.9$~G, 
which implies 
$\nu_{\rm m} (t^*)\approx 2.5 (\epsilon_e/0.1)^2 \times 10^{-2}$ GHz and $\nu_c(t^*)\ \approx 2.6 \times 10^{2}$ GHz. 
It follows that for young radio SNe $\nu_{\rm m} < \nu_c$ and the only relevant spectral break frequency that typically enters the $0.1-100$ GHz radio spectrum is $\nu_{\rm sa}$.

$\nu_{\rm{sa}}$ divides the radio spectrum into optically ``thick'' ($F_{\nu, \rm{thick}}$, for $\nu<\nu_{\rm sa}$), and optically ``thin"  ($F_{\nu, \rm{thin}}$, for  $\nu > \nu_{\rm sa}$). At frequencies below $\nu_{\rm sa}$, the optical depth to synchrotron emission is $\tau_{\rm sa}(\nu_{\rm sa })>1$ and in the absence of any other absorption mechanism SSA dominates \citep{RyLi}.
In the spectral regime of interest, $F_{\nu, \rm{thick}} \propto \nu^{5/2}$ for $ \nu_{\rm m} < \nu < \nu_{\rm sa}$. We note for completeness that at lower frequencies such that  $ \nu < \nu_{\rm m} < \nu_{\rm sa}$,
$F_{\nu, \rm{thick}} \propto \nu^{2}$ \citep[e.g.,][]{granot2002}.

In order to derive physical quantities from a single SED, we follow \citet{RyLi} and C98. The specific intensity $I_{\nu}$ can be expressed as follows:

\begin{equation}
\label{eq:I}
    I_\nu = S_\nu(\nu_{\rm sa})J_\nu(y,p),
\end{equation}
\noindent where
\begin{equation}
\label{eq:S}
    S_\nu(\nu_{\rm sa})=\frac{c_5}{c_6}(B\sin{\theta})^{-1/2} \Big(\frac{\nu_{\rm sa}}{2c_1}\Big)^{5/2},
\end{equation}

\begin{equation}
\label{Eq:nusa}
    \nu_{\rm sa} = 2c_1
    \bigg[\frac{4}{3}fRc_6\bigg]^{2/(p+4)}N_0^{2/(p+4)}(B\sin\theta)^{(p+2)/(p+4)},
\end{equation}
\begin{equation}
\label{eq:J}
    J_\nu(y,p)=y^{5/2} \Big(1-e^{-y^{-(p+4)/2}}\Big),
\end{equation}

\noindent
and 
$y \equiv \frac{\nu}{\nu_{\rm sa}}$.
The constant $c_1$ and the parameters $c_5(p)$, and $c_6(p)$ are defined in \citet{P1970} and provided in full in Appendix \ref{App:constants}. As these parameters are functions of $p$, it is typical in the radio SN literature to approximate $c_5\sim c_5(p=3)$ and $c_6\sim c_6(p=3)$, irrespective of what the best-fitting $p$ is for the SED. 
In this work we self-consistently account for the dependency of $c_5$ and $c_6$ on the electron power-law index $p$.
Additionally, we leave $B \sin{\theta}$  explicit in our derivation (as opposed to approximating $B \sin{\theta} \approx B$), as further physical quantities will not maintain the same exponential relationships between $B$ and $\sin\theta$.\footnote{See \citet{yang2018} for a recent discussion regarding the flux density dependence on the pitch angle $\theta$ distribution of the radiating electrons.} 
In Equations \ref{eq:I}, \ref{eq:S}, and \ref{eq:J} above, the radio emitting region is assumed to be a disk in the sky of radius $R$, thickness $s$, and volume $V=\pi R^2 s$. The filling factor $f$ is the fractional volume of a sphere of radius $R$ such that the emitting volume is
\begin{equation}
\label{Eq:volume}
    V= \pi R^2 s\equiv f\times 4/3 \pi R^3. 
\end{equation}
Canonically, a value $f=0.5$ is adopted, but we leave $f$ explicit in this work. 
The number density distribution of relativistic electrons per unit volume in energy space can be expressed as $dN/dE=N_0 E^{-p}$, with $N_0$ as the normalization. 
This relates to $K_0$ through $K_0 \gamma_e^{-p} d\gamma_e = N_0 E^{-p} dE$ and $E=\gamma_e m_e c^2$.
Following Equation \ref{Eq:Eedef}, $E_e=\int_{E_l}^{\infty} E\,N_0 E^{-p}\,{dE} = \int_{\gamma_m}^{\infty} (\gamma_e m_e c^2) K_0 \gamma_e^{-p} d\gamma_e$ and $E_l$ is the lowest energy of the electrons accelerated to a power law, $E_l= \gamma_m m_e c^2$. 
For $p>2$, the energy density in relativistic electrons of Equation \ref{Eq:Eedef} can be also expressed as 

\begin{equation}
    \label{eq:numdens_e}
    E_e=(N_0 E_l^{2-p})/(p-2).
\end{equation}
\noindent

We may now express the way Equation \ref{eq:numdens_e} relates to B:

\begin{equation}
    \label{eq:numdens_b}
    \frac{1}{\epsilon_B}\frac{B^2}{8\pi}=\frac{1}{\epsilon_e}\frac{N_0E_l^{(2-p)}}{(p-2)} . 
\end{equation}

The flux density $F_\nu$ of a source of radial extent $R$ and constant $I_\nu$ over the subtended solid angle, $\Omega = \pi R^2/D^2$  is $F_{\nu}= I_{\nu}\pi R^2/D^2$, where $D$ is the distance of the source.

Considering the asymptotic limits   $\nu \gg \nu_{\rm sa}$ (optically thick) and $\nu \ll \nu_{\rm sa}$ (optically thin), and using the equations above, C98 derives:
\begin{equation}
    \label{eq:thick}
    F_{\nu, \rm{thick}}(\nu) = \frac{c_5}{c_6}(B\sin\theta)^{-\frac{1}{2}}\Big(\frac{\nu}{2c_1}\Big)^{\frac{5}{2}}\frac{\pi R^2}{D^2},
\end{equation}
\noindent which scales as $F_{\nu, \rm{thick}} \propto \nu^{5/2}$, and the optically thin flux density $F_{\nu, \rm{thin}} \propto \nu^{-(p-1)/2}$ is:
\begin{equation}
    \label{eq:thin}
    F_{\nu, \rm{thin}} (\nu) =  c_5 \Big(\frac{\nu}{2c_1}\Big)^{\frac{-(p-1)}{2}}(B\sin \theta)^{\frac{p+1}{2}}N_0 \frac{4}{3} \pi f \frac{R^3}{D^2}.
\end{equation}

\noindent
We define a break frequency parameter $\nu_{\rm{brk}}$ as the spectral frequency at which the asymptotically thick and thin flux densities (Equations \ref{eq:thick} and \ref{eq:thin}) meet: $F_{\nu,\rm{thin}} (\nu_{\rm{brk}}) = F_{\nu,\rm{thick}} (\nu_{\rm{brk}})\equiv F_{\rm{brk}}$. 
After combining Equations \ref{eq:numdens_b}, \ref{eq:thick}, and \ref{eq:thin} and solving for $B$ and $R$, we find the results below. We provide coefficients containing several digits to avoid rounding errors if a reader were to use these equations as provided.

\begin{dmath}
\label{eq:B}
    B = (2.50 \times10^9)
    \Big(\frac{\nu_{\rm{brk}}}{\rm{5\,GHz}} \Big) 
    \Big(\frac{1}{c_1}\Big) \\
    \Bigg[
    \frac{
    4.69\times10^{-23}
    (\frac{E_l}{\rm{erg}})^{2(2-p)}\epsilon_B^2 c_5 \sin(\theta)^{\frac{1}{2}(-5-2p)}
    }
    {
    (p-2)^2
    (\frac{D}{\rm{Mpc}})^2\epsilon_e^2 (\frac{f}{0.5})^2 
    (\frac{F_{\rm{brk}}}{\rm{Jy}})c_6^3
    }\Bigg]^{\frac{2}{13+2p}} \rm{G},
\end{dmath}

\begin{dmath}
\label{eq:R}
R = 
(2.50 \times10^9)^{-1}
c_1
\bigg(\frac{\nu_{\rm{brk}}}{\rm{5\,GHz}}\bigg)^{-1}
\bigg[
12  
\epsilon_B
c_5^{-(6 + p)}
c_6^{(5 + p)} \\
(9.52\times10^{25})^{(6+p)}
\sin^2{\theta} 
\pi^{-(5 + p)} 
\bigg(\frac{D}{\rm{Mpc}}\bigg)^{2 (6 + p)} \\
\bigg(\frac{E_l}{\rm{erg}}\bigg)^{(2 - p)} 
\bigg(\frac{F_{\rm{brk}}}{\rm{Jy}}\bigg)^{(6 + p)} \\
\bigg(
\epsilon_e
(p - 2)
\bigg(\frac{f}{0.5}\bigg)
\bigg)^{-1}
\bigg]^{1/(13 + 2 p)}
\rm{cm}.
\end{dmath}

For direct comparison to C98, we evaluate $B$ and $R$ at typical values of the physical parameters for radio SNe, i.e., 
$p = 3$, $E_l=m_ec^2$, and $\sin \theta \approx 1$
\citep[e.g.,][]{weiler02, Berger02, soderberg03bg, 
chevandfrans, Corsi14}. 
Below, we demonstrate that Equation \ref{eq:B} reduces to Equation 12 in C98; and Equation \ref{eq:R} reduces to their Equation 11: 

\begin{dmath}
\label{eq:chevb}
B = 0.5761 
\bigg(\frac{\nu_{\rm{brk}}}{\rm{5\,GHz}}\bigg) \\
\bigg[
\epsilon_B^{-1} 
\bigg(\frac{D}{\rm{Mpc}}\bigg)
\epsilon_e
\bigg(\frac{f}{0.5}\bigg)
\bigg(\frac{F_{\rm{brk}}}{\rm{Jy}}\bigg)^{1/2}
\bigg]^{-4/19} \rm{ G},
\end{dmath}

\begin{dmath}
\label{eq:chevr}
R = 8.759\times10^{15} 
\bigg(\frac{\nu_{\rm{brk}}}{\rm{5\,GHz}}\bigg)^{-1} \\
\bigg[
\bigg(\frac{D}{\rm{Mpc}}\bigg)^{18} 
\epsilon_B 
\bigg(\frac{F_{\rm{brk}}}{\rm{Jy}}\bigg)^9
  \epsilon_e^{-1} 
  \bigg(\frac{f}{\rm{0.5}}\bigg)^{-1}
  \bigg]^{1/19} \rm{cm}.
\end{dmath}

From Equations \ref{Eq:EBdef}, \ref{Eq:volume}, \ref{eq:B}, and \ref{eq:R}, the postshock energy corresponding to the energy density  $\frac{2}{1+i}\rho_{\rm{CSM}}v^2$ is $U\equiv (\frac{2}{1+i} \rho_{\rm{CSM}}v^2) \times V =B^2/(\epsilon_B 8 \pi)\times ( 4/3\pi R^3f)$, or:

\begin{dmath}
\label{eq:U}
U = 
 c_1
  (3.33\times10^{-11})
   \epsilon_B^{-1}
10^{\frac{75 (6 + p)}{13 + 2 p}} \\
\bigg(\frac{f}{0.5}\bigg)
\bigg(\frac{\nu_{\rm{brk}}}{\rm{5\,GHz}}\bigg)^{-1}
 \Bigg[
 3.086^{6(6+p)}
 4.411\times10^{-96} \\
\bigg(\frac{D}{\rm{Mpc}}\bigg)^{28 + 6 p}
\bigg(\frac{F_{\rm{brk}}}{\rm{Jy}}\bigg)^{14 + 3 p}
 \pi^{-3 (1 + p)}
 \sin{\theta}^{-4 (1 + p)} \\
 c_5^{-(14 + 3 p)}
 c_6^{3 (1 + p)} 
 \bigg(
  2
   \epsilon_B
 \epsilon_e^{-1}
  (p - 2)^{-1} \\
\bigg(\frac{E_l}{\rm{erg}}\bigg)^{2 - p}
 \bigg(\frac{f}{0.5}\bigg)^{-1}
 \bigg)^{11}
 \bigg]^{1/(13 + 2p)}
 \rm{erg.}
\end{dmath}
We note that for the case of a monatomic gas and a strong shock, the post-shock thermal energy is $U_{th}= \frac{9}{8} V \rho_{\rm{CSM}}v^2$ (e.g., \citealt{petropoulou16}, see Appendix \ref{App:microphysics} for a comparison). 

Substituting $p = 3$, $E_l=m_ec^2$, and $\sin \theta \approx 1$ as above, 
our Equation \ref{eq:hoU} reduces to Equation 12 in \citet{Ho19}:
\begin{dmath}
\label{eq:hoU}
U = 1.859\times10^{46} 
\epsilon_B^{-1} 
\bigg(\frac{\nu_{\rm{brk}}}{\rm{5\,GHz}}\bigg)^{-1} \\
\bigg[\epsilon_B^{11}
  \epsilon_e^{-11} 
  \bigg(\frac{f}{0.5}\bigg)^{8} 
  \bigg(\frac{D}{\rm{Mpc}}\bigg)^{46}
  \bigg(\frac{F_{\rm{brk}}}{\rm{Jy}}\bigg)^{23} \bigg]^{1/19}  \rm{erg}.
\end{dmath}

\subsection{Radio Synchrotron Emission in SNe: environment density $\rho_{\rm{CSM}}$ and pre-explosion mass-loss rate $\dot M$} 
\label{SubSec:Mdot}
A physical quantity of interest is the density of matter in the pre-shocked circumstellar medium, which directly traces the mass-loss history of the progenitor system before the star's death.  
We derive the density of the unshocked CSM, $\rho_{\rm CSM}$, by rearranging Equation \ref{Eq:EBdef} and substituting for $B$ and $R$ (Equations \ref{eq:B} and \ref{eq:R}). 
Additionally, we follow C98 in assuming the evolution of the shock radius with time follows a power law $R\propto t^q$ (either globally or locally), such that $v\equiv dR/dt = q R/t$ (see Appendix \ref{App:microphysics}, and $R(t)$ in Figure \ref{fig:brv} for an example). Thus,

\begin{dmath}
\label{eq:rho}
\rho_{\rm{CSM}} = 5.8\times10^{45} 
\bigg(\frac{\nu_{\rm{brk}}}{\rm{5GHz}}\bigg)^4 
\bigg(\frac{t}{\rm{days}}\bigg)^2 
(i+1) \\
\bigg(
c_1^4 
\epsilon_B 
q^2\bigg)^{-1}
\bigg[
3.355
\times10^{-92}
\times 10^{-50(6+p)}  \\
\times 3.086^{-4(6+p)}
c_5^{2(8+p)} 
c_6^{-2(11+p)} \\
\bigg(\frac{D}{\rm{Mpc}}\bigg)^{-4(8+p)}
\bigg(\frac{F_{\rm{brk}}}{\rm{Jy}}\bigg)^{-2(8+p)}   
\pi^{2(5+p)}
\sin(\theta)^{-2(7+2p)}
\\ \bigg[
\bigg(\frac{E_l}{\rm{erg}}\bigg)^{(p-2)}
\epsilon_B^{-1}
\epsilon_e  
\bigg(\frac{f}{0.5}\bigg)
(p-2)
\bigg]^{-6}
\bigg]^{\frac{1}{(13+2p)}} \rm{g\, cm^{-3}}.
\end{dmath}

For a shock that originates from the interaction between the outer layers of dense SN ejecta (which has a power-law distribution of density in radius $\rho_{\rm ej}\propto R^{-n}$) and a power-law density distribution of CSM ($\rho_{\rm{CSM}}\propto R^{-s}$), the shock radius evolution is described by the self-similar solutions in \citet{chev82}.\footnote{The power-law density of the unshocked medium, $s$, is not to be confused with the model smoothing parameter $s$ in Section \ref{sec:fit}. 
We use this notation for ease of comparison to \citet{chev82}.} $R(t)\propto t^{\frac{n-3}{n-s}}$, which implies  $q=\frac{n-3}{n-s}$. 
We note this relationship only holds in the regime of the self-similar solution, i.e., so long as $n>5$ and $s<3$ \citep{chev82}. 
For a core-collapse SN originating from a compact, massive, stripped progenitor such as a Wolf-Rayet (WR) star $n\approx 10$, and the CSM created from massive stellar winds  scales with $s=2$, which leads to a value of $q=7/8$ that is typically applied in the literature \citep[see e.g.,][]{chevandfrans}.  
From Equation \ref{eq:rho}, and a radio spectrum, the inferred pre-shock CSM density scales as
$\rho_{\rm{CSM}}\propto \frac{1+i}{q^2}$.   

Introducing the mean molecular weight per electron parameter $\mu_e$ and proton mass $m_p$ for the CSM, the electron number density in the unshocked CSM is $n_e= \frac{\rho_{\rm{CSM}}}{m_p\mu_e}$, or:

\begin{dmath}
    \label{eq:ne}
    n_e =
    5.8\times10^{45} 
    (i+1)
    \bigg(\frac{\nu_{\rm{brk}}}{\rm{5GHz}}\bigg)^4 
    \bigg(\frac{t}{\rm{days}}\bigg)^2 \\
    (c_1^4 
    \epsilon_B 
    m_p 
    \mu_e
    q^2)^{-1} \\
    \bigg[3.355
    \times10^{-92}
    \times 10^{-50(6+p)}
    \times3.086^{-4(6+p)}
    c_5^{2(8+p)} 
    c_6^{-2(11+p)} \\
    \bigg(\frac{D}{\rm{Mpc}}\bigg)^{-4(8+p)}
    \bigg(\frac{F_{\rm{brk}}}{\rm{Jy}}\bigg)^{-2(8+p)}
    \pi^{2(5+p)}
    \sin(\theta)^{-2(7+2p)} \\
    \bigg[\bigg(\frac{E_l}{\rm{erg}}\bigg)^{(p-2)} 
    \epsilon_B^{-1} 
    \epsilon_e 
    \bigg(\frac{f}{0.5}\bigg)
    (p-2)\bigg]^{-6}\bigg]^{1/(13+2p)} \rm{cm^{-3}}.
\end{dmath}

During its lifetime, a star loses material to its surroundings and sits in a polluted environment of its own debris. 
For stars that lose mass through a constant mass-loss rate $\dot M$ and wind velocity  $v_{w}$, the CSM density is related to $\dot M$ as $\rho_{\rm{CSM}}=\frac{\dot M}{4\pi R^2 v_{w}}$. 
Combining Equations \ref{eq:R} and \ref{eq:rho}, the mass-loss rate at a radius probed by the forward shock thus satisfies the relation:

\begin{dmath}
\label{eq:ML}
    \frac{\dot{M}}{v_{w}} \bigg(\frac{1000\, \kms}{10^{-4}M_{\sun}\rm{yr}^{-1}}\bigg) =
1.850 \times10^{14} 
\bigg(\frac{\nu_{\rm{brk}}}{\rm{5GHz}}\bigg)^2 
\bigg(\frac{t}{\rm{days}}\bigg)^2 
(i+1)\\
  (c_1^2 \epsilon_B q^2)^{-1}
 \\
\bigg[
\frac{
(4.688\times10^{-23})
  \big(\frac{E_l}{\rm{erg}}\big)^{2(2-p)}
  \epsilon_B^2 
  c_5 
  }{
    \sin{\theta}^{(5+2p)/2}
  \big(\frac{D}{\rm{Mpc}}\big)^2 
  \epsilon_e^2 
 \big(\frac{f}{0.5}\big)^2 
\big(\frac{F_{\rm{brk}}}{\rm{Jy}}\big)
  (p-2)^2 
  c_6^3}
  \bigg]
  ^{\frac{4}{(13+2p)}}.
\end{dmath}

Similarly to $\rho_{\rm{CSM}}$, both $n_{e} $ and $\dot M$ scale as $\propto \frac{i+1}{q^2}$. 
We emphasize that radio observations constrain the ratio $\dot M/v_{w}$ by providing $\rho_{\rm CSM}$ and that an independent estimate of $v_{w}$, from optical spectroscopy for example, is needed to resolve for the model degeneracy.
For CSM environments that deviate from a wind profile, Equation \ref{eq:ML} provides an estimate of the \emph{effective} mass-loss rate in units of $v_{w}$.

When substituting $q=7/8$, $p=3$, $i=1$, $E_l=m c^2$ and $\sin \theta \approx 1$, the expression of Equation \ref{eq:ML} can be compared to Equation 23 in \citet{chevandfrans}:

\begin{dmath}
\frac{\dot{M}}{v_{w}} \bigg(\frac{1000 \kms}{10^{-4}M_{\sun}\rm{yr}^{-1}}\bigg) \setq
2.5 \times 10^{-5}
\bigg(\frac{\nu_{\rm{brk}}}{5\rm{GHz}}\bigg )^2 
\bigg( \frac{t}{\rm{days}} \bigg)^2
\epsilon_B^{-1} \\
\Bigg[
\epsilon_B^{2}
\bigg( \frac{D}{\rm{Mpc}} \bigg)^{-2}
\bigg( \frac{F_{\rm{brk}}}{\rm{Jy}} \bigg)^{-1}
\epsilon_e^{-2}
\bigg( \frac{f}{0.5} \bigg)^{-2}
\bigg]^{4/19}.
\end{dmath}
When we evaluate our Equation \ref{eq:ne} instead at $q=1$ and $i=5/3$ and assume the material is fully ionized hydrogen (such that $\mu_e=1$), 
we recover Equation 16 of \citet{Ho19} (we note that they write their expression in terms of luminosity):

\begin{dmath}
\label{ne_ho}
n_e \seti 
1.02418
 \bigg(\frac{\nu_{\rm{brk}}}{\rm{5GHz}}\bigg)^4 
  \bigg(\frac{t}{\rm{days}}\bigg)^2 
  \epsilon_B^{-1}\\
\bigg[
\epsilon_B^{-3} 
 \bigg(\frac{D}{\rm{Mpc}}\bigg)^{22} 
 \epsilon_e^{3} 
  \bigg(\frac{f}{0.5}\bigg)^{3} 
   \bigg(\frac{F_{\rm{brk}}}{\rm{Jy}}\bigg)^{11}
   \bigg]^{-2/19} \rm{cm^{-3}}.
\end{dmath}

Equation \ref{eq:ML} reduces to:

\begin{dmath}
\label{Eq:masslossHo}
\frac{\dot{M}}{v_{w}} \bigg(\frac{1000\, \kms}{10^{-4}M_{\sun}\rm{yr}^{-1}}\bigg) \seti
2.6 \times 10^{-5}\,
\bigg( \frac{\nu_{\rm{brk}}}{5\rm{GHz}} \bigg)^2
\bigg( \frac{t}{\rm{days}} \bigg)^2
\epsilon_B^{-1} \\
\Bigg[
\epsilon_B^{2}
\bigg( \frac{D}{\rm{Mpc}} \bigg) ^{-2}
\bigg( \frac{F_{\rm{brk}}}{\rm{Jy}} \bigg)^{-1}
\epsilon_e^{-2}
\bigg( \frac{f}{0.5} \bigg)^{-2}
\bigg]^{4/19},
\end{dmath}
which is Equation 23 in \citet{Ho19} (in terms of flux and distance, rather than luminosity) once one accounts for the facts that: (i) Equation 23  in \citet{Ho19} reports the wrong scaling for the filling factor $f$; (ii) the reported normalization is also not correct ($\gtrsim 10$ times larger than the correct value). See also footnote 4 of \citet{Yao21}. Our Equation \ref{Eq:masslossHo} above gives the correct parameter scalings and normalization.

We conclude with considerations regarding the impact different values of $i$ and $q$ will have on pre-shock CSM density and mass-loss rates.
In the radio SN literature, a diversity of parameterization choices map to values of $i$ and $q$ that we explore in Appendix \ref{App:microphysics}. 
It follows that  inferred $\rho_{\rm{CSM}}$ values are \emph{not} directly comparable to one another, even when the same microphysical parameters $\epsilon_e$ and $\epsilon_B$ are chosen. 
For example, using  $i=5/3$ and $q=1$ (as in \citealt{Ho19}) leads to a factor of $\frac{i+1}{q^2}=8/3\approx2.67$ in $\rho_{\rm{CSM}}$; and \citet{chevandfrans} choose $i=1$ and $q=7/8$ which leads to $\frac{1+i}{q^2}\approx2.61$. Similarly, Equation \ref{Eq:epsilonBMatsuoka} employed by \citet{matsuoka2020} is equivalent to $i=-1/2$ and $q=1$; this definition leads to $\frac{1+i}{q^2}=1/2$. 
Amongst a range of assumed $i$ and $q$ values in the literature (Appendix \ref{App:microphysics}), the ratio  $\frac{1+i}{q^2}$ spans values 1/2 to 8/3, leading to reported values of $\rho_{\rm{CSM}}$ (as well as $n_e$ and $\dot M$) that are \textit{systematically different up to a factor $\approx5$.}
We thus caution the reader regarding direct comparisons of $\dot M$ (or $n_e$ or $\rho_{\rm{CSM}}$) values from independent works, even when the same shock microphysical parameters are used.

\subsection{Radio SED fitting}
\label{sec:fit}
In this section we describe the fitting procedure used on the radio SEDs to extract the physical parameters of the emitting source based on the physics described in the previous sections.  
We model the radio SED produced by synchrotron emission described in \S\ref{SubSec:SynchRBU}  with a smoothed  broken power law (BPL) of the form:

 \begin{equation}
    \label{eq:bpl}
      F_{\nu} = F_{\rm{brk}}\bigg[
      \bigg(\frac{\nu}{\nu_{\rm{brk}}}\bigg)^{\alpha_1/s}
      +
      \bigg(\frac{\nu}{\nu_{\rm{brk}}}\bigg)^
      {\alpha_2/s}
      \bigg]^s.
 \end{equation}

This functional form is widely used in the radio SN and non-jetted TDE literature (e.g., \citealt{soderberg03bg}, \citealt{Soderberg06c}; \citealt{Alexander20} and references therein). 
The smoothing parameter $s$ represents how abruptly the SED turns over from the ``optically thick" slope ($\alpha_2$) to the ``optically thin" slope ($\alpha_1$). 
For an SSA spectrum with $\nu_{\rm m}<\nu_{\rm sa}$ $\alpha_1=-(p-1)/2$ and we expect $\alpha_2=5/2$ (Equations \ref{eq:thick} and \ref{eq:thin}). 
Sharper transitions between slopes are associated with smaller $s$ values.
For the spectral regime of interest, $\nu_{\rm{brk}}$ can be identified as $ \nu_{\rm sa}$.  $F_{\rm{brk}}$ represents the peak flux density where the asymptotic power-laws describing the optically thin and the optically thick spectrum meet (i.e., the $F_{\rm{brk}}$ parameter in Equation \ref{eq:bpl} is the  break flux density that appears in the equations of the previous sections).

In general, $\nu_{\rm{brk}}\equiv \nu_{\rm{brk}}(t) $ and $F_{\rm{brk}}\equiv F_{\rm{brk}}(t)$. For a radio SN, the typical behavior is to observe a migration of  $\nu_{\rm{brk}}$ to lower frequencies as time progresses and the emitting region expands over a larger volume, leaving a broader range of higher frequencies optically thin to synchrotron radiation. 
We fit each SED without imposing a pre-defined temporal evolution of the parameters (in the SN literature a power-law evolution of $\nu_{\rm{brk}}(t)$ and  $F_{\rm{brk}}(t)$ is typically assumed, \citep[e.g.,][]{soderberg03bg})
allowing us to reconstruct the actual density profile of the CSM created by the star before explosion, and hence the true mass-loss history of the progenitor. An example of this procedure on our data is represented in Figures \ref{fig:SED_evol} and \ref{fig:freqslices}, and the full SED evolution is Figure \ref{fig:SED} in the Appendix.

 The SED power-law slopes $\alpha_1$ and $\alpha_2$ are free parameters that can, in principle, take different values in each of the SEDs.
Our fitting procedure provides the flexibility to jointly fit all the available SEDs with common values of $\alpha_1$ and $\alpha_2$, along with the smoothing parameter $s$.
At late times when the peak frequency has migrated below our observable frequency range, we fit the SED to a single-sided power law and include this information in a joint-fit of the slopes. The information from single-sided SEDs provide lower and upper limits on  physical parameters. 
 Our fitting procedure thus does not assume any value for $p$; instead $p$ can be derived from the best-fitting optically thin slope $\alpha_1$. We use the BPL function of Equation \ref{eq:bpl} with \texttt{lmfit} and perform a least squares minimization.

We point out that an important source of confusion in the radio SN literature is that the $F_{\rm{brk}}$ of Equations \ref{eq:B}${-}$\ref{eq:ne} is often incorrectly identified as the peak of the smoothed broken power-law flux density, as opposed to the intersection of the two asymptotic power-laws.
This is important because this implies that the true $F_{\rm{brk}}$ is underestimated by a factor of $2^s$  that propagates into all estimates of the physical parameters (see Equations 
\ref{eq:B}${-}$\ref{eq:ne}).

\section{Modeling of the radio SN\,2004C and Physical inferences} \label{sec:phys}

\subsection{Radio SED fitting of SN\,2004C} \label{sec:model}
The radio data of \sn\  are reported in Table \ref{tab:data}. We consider data acquired within date ranges $\delta t/t<0.025$ as part of the same SED. This procedure identifies 28 individual radio SEDs spanning the range $\delta t= 41.41-1938.22$ days since explosion.
We modeled each SED with the BPL function of Equation \ref{eq:bpl}, and we jointly fit all the SEDs with the same $\alpha_1$, $\alpha_2$ and $s$ value, while allowing $F_{\rm{brk}}$ and $\nu_{\rm{brk}}$ to evolve from one SED to the other. We explored with our software four cases: (i) fixing both the thick slope $\alpha_2 = 5/2$ and $s=-1$ (``BPL1"), (ii) fixing only $s=-1$ (``BPL2"), (iii) fixing only the thick slope $\alpha_2 = 5/2$ (``BPL3"), and (iv) allowing both slopes and $s$ vary (``BPL4"). 

We compare these models using an F-Test and report our results in Appendix \ref{app:ftest}, Table \ref{tab:models}. 
We adopt the standard $p$-value threshold of 5\% to reject a given model, and $\chi^2$ when the degrees of freedom are equal. We identify BPL1 (\fitcase) as the favored model by these statistics. 
The best-fitting model parameters for BPL1 are reported in Table \ref{tab:fit}. 
Figures \ref{fig:SED_evol}, \ref{fig:freqslices}, and \ref{fig:SED} show the best fitting model and the data.

\begin{deluxetable}{|c|c|c|c|}[h!]
\label{tab:fit}
\tablecaption{SED best-fitting parameters for our baseline model (BPL1), which has an assumed smoothing parameter and optically thick slope \fitcase\, and an 
optically thin slope $\alpha_1 = $ \thin . ``N/A'' entries show SEDs containing only one point and could not be fit to a one-sided power law.}
\tablecolumns{4}
\tablewidth{\textwidth}
\tablehead{
\thead{SED \\ Number} &
\thead{Days Since \\ Explosion} &
\thead{$\nu_{\rm{brk}}$ (GHz)} &
\thead{$F_{\rm{brk}}$ ($\mu$Jy)} 
}
\startdata
0 & 39.65 & N/A & N/A \\ \hline
1 & 41.41 -- 42.36 & 9.8 $\pm$ 0.6 & 3710 $\pm$ 290 \\ \hline
2 & 55.36 & 9.0 $\pm$ 0.5 & 4320 $\pm$ 300 \\ \hline
3 & 82.40 & 10 $\pm$ 0.6 & 11400 $\pm$ 700 \\ \hline
4 & 94.15 & 10 $\pm$ 0.4 & 14300 $\pm$ 700 \\ \hline
5 & 101.21 &  9.3 $\pm$ 0.4 & 16000 $\pm$ 800 \\ \hline
6 & 114.07 &  8.7 $\pm$ 0.5 & 18900 $\pm$ 1400 \\ \hline
7 & 120.14 & 9.2 $\pm$ 0.4 & 18000 $\pm$ 800 \\ \hline
8 & 126.08 & 9.3 $\pm$ 0.4 & 18400 $\pm$ 900 \\ \hline
9 & 140.14 & 8.6 $\pm$ 0.3 & 18600 $\pm$ 800 \\ \hline
10 & 150.18 & 8.0 $\pm$ 0.3 & 19400 $\pm$ 900 \\ \hline
11 & 167.16 & 8.3 $\pm$ 0.6 & 16000 $\pm$ 800 \\ \hline
12 & 178.98 -- 179.04 & 6.4 $\pm$ 0.3 & 14800 $\pm$ 700 \\ \hline
13 & 196.08 & 5.2 $\pm$ 0.5 & 16900 $\pm$ 1000 \\ \hline
14 & 202.05 & 6.6 $\pm$ 0.9 & 13700 $\pm$ 1000 \\ \hline
15 & 224.98 -- 225.99 & 4.4 $\pm$ 0.5 & 14900 $\pm$ 1000 \\ \hline
16 & 241.97 -- 242.02 & 4.9 $\pm$ 0.4 & 13200 $\pm$ 800 \\  \hline
17 & 284.84 & $\leq$ 4.9 & $\geq 9400$ \\ \hline
18 & 304.59 & $\leq$ 4.9 & $\geq 11000$  \\ \hline
19 & 343.67 & $\leq$ 4.9 &  $\geq 8500$  \\ \hline
20 & 406.20 -- 416.27 & $\leq$ 4.9 & $\geq  9000$ \\ \hline
21 & 523.47 -- 523.90 & $\leq$ 4.9 & $\geq 9000$ \\ \hline
22 & 615.94 & $\leq$ 4.9 & $\geq 7500 $ \\ \hline
23 & 701.60 & $\leq$ 4.9 & $\geq 11000 $ \\ \hline
24 & 813.41 & $\leq$ 4.9  & $\geq 5900 $ \\ \hline
25 & 1260.10 & $\leq$ 4.9 & $\geq 3100 $  \\ \hline
26 & 1378.74 & $\leq$ 4.9 & $\geq 2000 $ \\ \hline
27 & 1785.5 & $\leq$ 4.9 & $\geq 1500 $ \\ \hline
28 & 1938.22 & $\leq$ 4.9 & $\geq 1500 $ \\ \hline
29 & 2793.65 & N/A & N/A
\\ \hline
\enddata

\end{deluxetable}

\subsection{Inferred physical parameters of SN\,2004C: $p$, $R(t)$, $B(R)$}
\label{SubSec:04CphysicsRB}
\begin{figure*}[!ht] \centering
\begin{tabular}{cc}
    \includegraphics[width=65mm]{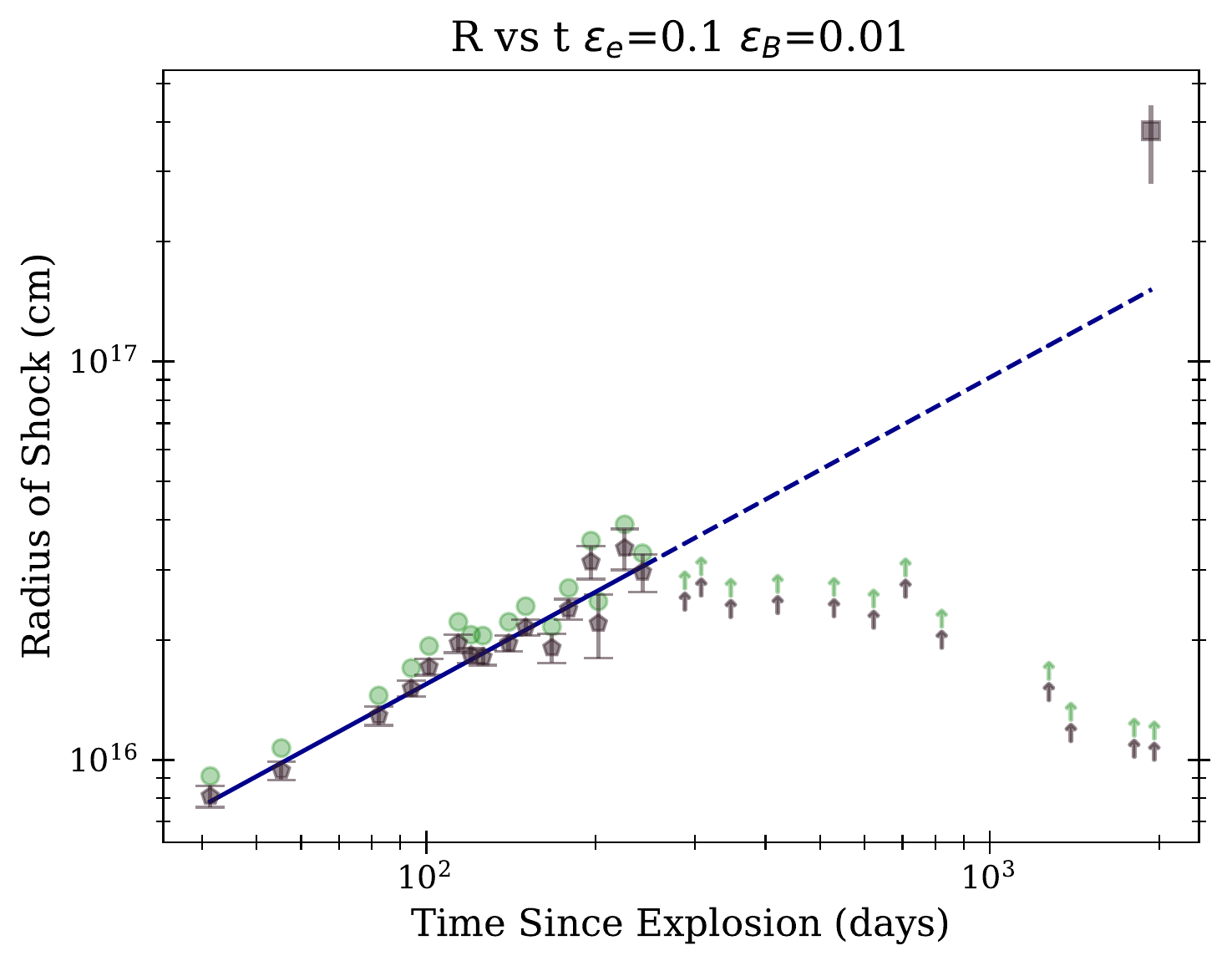} &
    \includegraphics[width=67mm]{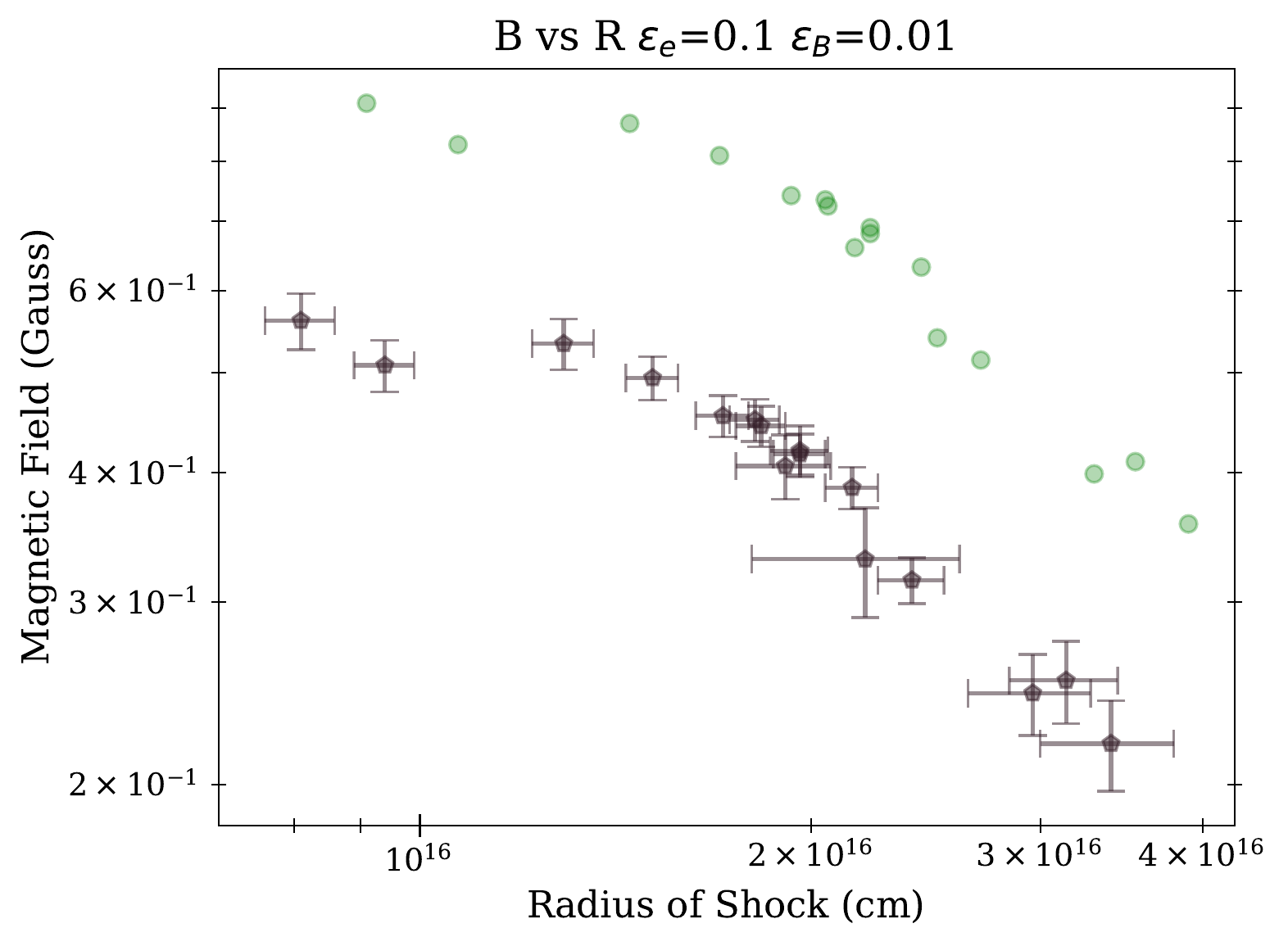} \\
    \includegraphics[width=65mm]{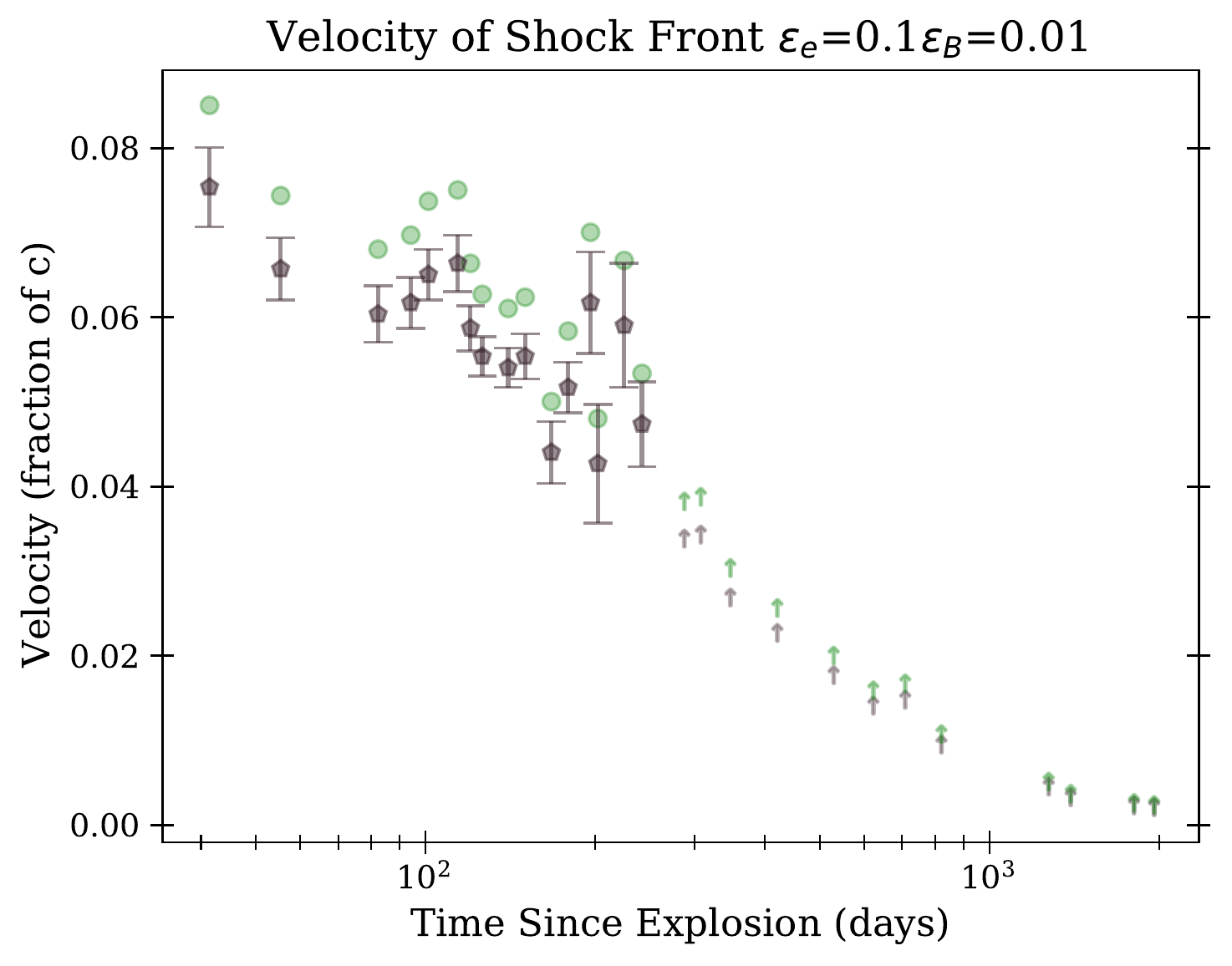} &
        \includegraphics[width=67mm]{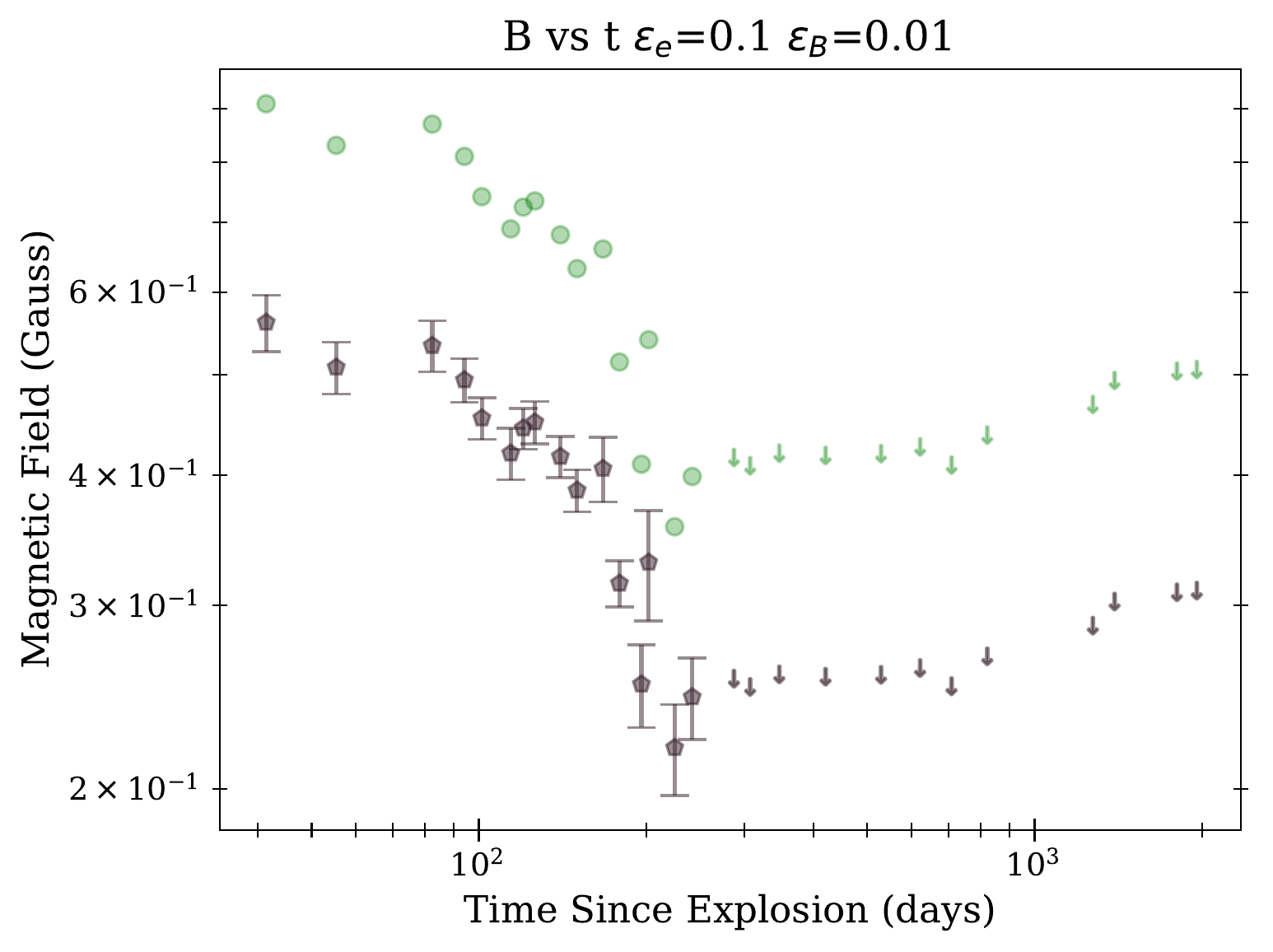} \\
\end{tabular}
\caption{Physical parameters derived from $F_{\rm{brk}}$ and $\nu_{\rm{brk}}$ using our re-derived Equations \ref{eq:B} and \ref{eq:R} for different choices of microphysical parameters. \textbf{Purple points} correspond to the favored broken power law model (\fitcase) with $\epsilon_B = 0.01, \epsilon_e =0.1$. \textbf{Green points} are in equipartition ($\epsilon_B = \epsilon_e =1/3$) and have the same uncertainties as the purple points. 
\textbf{Top, left panel; blue line:} Power-law fit $R\propto t^{q}$ of the purple points, where $q=\text{\q}$.
The dashed line is a continuation of this fit until the time of the VLBA measurement. 
\textbf{Top, left panel; square point:} Our VLBA measurement of the radius (\vlbaMeas\ cm) obtained at 1931 days since explosion. A re-fit incorporating this value agrees within uncertainties with the value inferred from modeling of VLA observations at $<300$ days since explosion.}
\label{fig:brv}
\end{figure*}

We use the best-fitting parameters of Table \ref{tab:fit}, their uncertainties, and covariance information to estimate the physical parameters of the shock launched by SN\,2004C and the circumstellar environment with which it interacts.
First, the optically thin slope $\alpha_1=-(p-1)/2$ implies a power-law index of the electron distribution $p=$\pvalue, in-line with typical values inferred from SN from the radio modeling of SNe $p\approx 3$ \citep[e.g.,][]{chevandfrans, soderberg03bg} and recently interpreted by \citet{Diesing21}. Values of $p$ show consistency amongst all our fit models, falling within $\sim0.1$ of one another.
Next, we estimate $R(t)$ and $B(t)$ using Equations \ref{eq:R} and \ref{eq:B}, respectively. 
In our analysis we adopt $i=1$ and we infer $q=$ \q\ (Figure \ref{fig:brv}). We present our results for the case of equipartition ($\epsilon_e=1/3$, $\epsilon_B=1/3$) and for our fiducial case  $\epsilon_e=0.1$ and $\epsilon_B=0.01$. 
We note that the radius and magnetic field scale as $R\propto \epsilon_e^{\frac{-1}{(13+2p)}}\epsilon_B^{\frac{1}{(13+2p)}}$ and $B\propto \epsilon_e^{\frac{-4}{(13+2p)}} \epsilon_B^{\frac{4}{(13+2p)}}$ and are thus minimally sensitive to specific assumptions of the shock microphysical parameters. 
Instead, $U(t)\propto \epsilon_e^{\frac{-11}{(13+2p)}}\epsilon_B^{\frac{11}{(13+2p)} -1}$, 
$\rho_{\rm{CSM}}(r)\propto \epsilon_e^{\frac{-6}{(13+2p)}}\epsilon_B^{\frac{6}{(13+2p)} - 1}$ and 
$\dot M (r)\propto \epsilon_e^{\frac{-8}{(13+2p)}}\epsilon_B^{\frac{8}{(13+2p)} - 1}$  are sensitive to the assumed  $\epsilon_e$ and $\epsilon_B$ values and the assumption of equipartition conveniently provides solid lower-limit on their values. 
Relatedly, radius scales as $R\propto D^{\frac{2(6+p)}{(13+2p)}}$ and mass loss as $\dot{M}\propto D^{\frac{-16}{(13+2p)}}.$ Changes to the distance by a factor $N$ moderately affect our results; for $p\approx3$, $R$ will change by a factor of $N^{0.9}$ and $\dot{M}$ by a factor of $N^{-0.8}$.

Figure \ref{fig:brv} shows the evolution of the forward shock radius and shock velocity with time, implying mild deceleration of the shock as it interacts with material in the ambient medium. 
We find $R(t)\propto$ $t^{\text{\q}}$ over the distance range \innerradius $-$ \outerradius cm using observations acquired $41 - 1938$ days since explosion. 
With the VLBA we inferred $R=3.8^{+0.6}_{-1.0}\times 10^{17}$ cm at 1931 days since explosion. 
The extrapolation of our best-fitting FS radius derived from VLA observations to the time of the VLBA observations is $R=$ \VLBApredR\ cm,  which is consistent, but slightly lower, than the interferometry measurement at the  $3\,\sigma$ level (Fig.\ \ref{fig:brv}). We remark that this is an extrapolation over almost one order of magnitude in time.
Including the VLBA measurement in the $R(t)$ fit we find  $q=$ \qVLBA, which is consistent within uncertainties with the radius expansion rate that we infer at $< 242$ days since explosion.
The forward shock has an average shock velocity of $\approx 0.06 c$ throughout the duration of our observations.  
For a power-law evolution of the shock radius with time, $R(t)\propto t^q$ the velocity is $v=q\frac{R}{t}$. 

Instead of assuming a specific progenitor type and CSM density profile to compute $q$ using the self-similar solutions by \citet{chev82}, we allow our code to be as agnostic and self-consistent as possible: we derive $q$ by taking the slope of $R(t)$ (solid line in Figure \ref{fig:brv}).
Other fitting methods that hold the SED slope constant for synchrotron emission or allow all BPL fit parameters to vary lead to reasonably similar values. Appendix \ref{App:microphysics} demonstrates the effect this has on derived physical quantities.

\begin{figure*}[!ht]\centering
{\includegraphics[width=.45\linewidth]{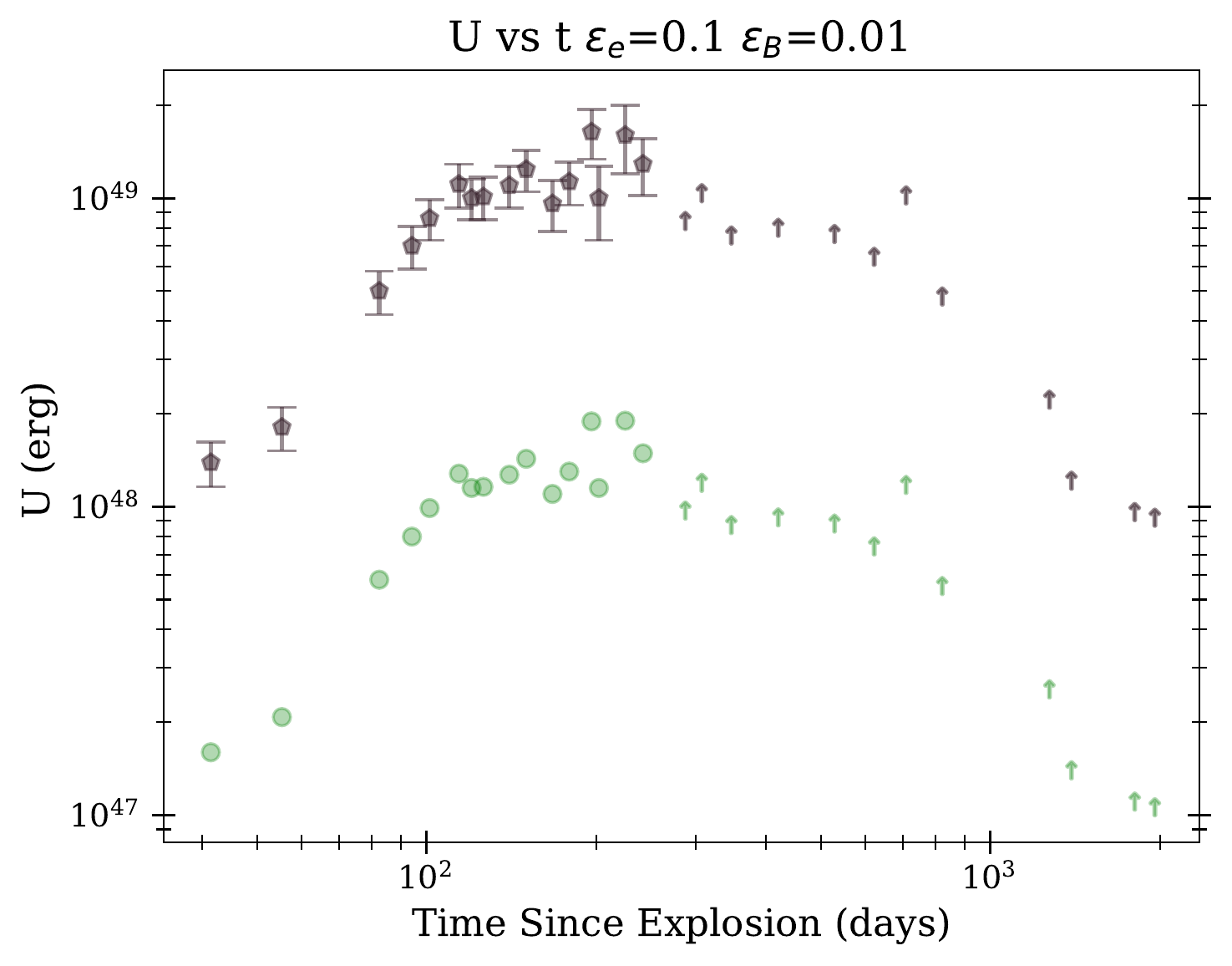}}\hfill
{\includegraphics[width=.45\linewidth]{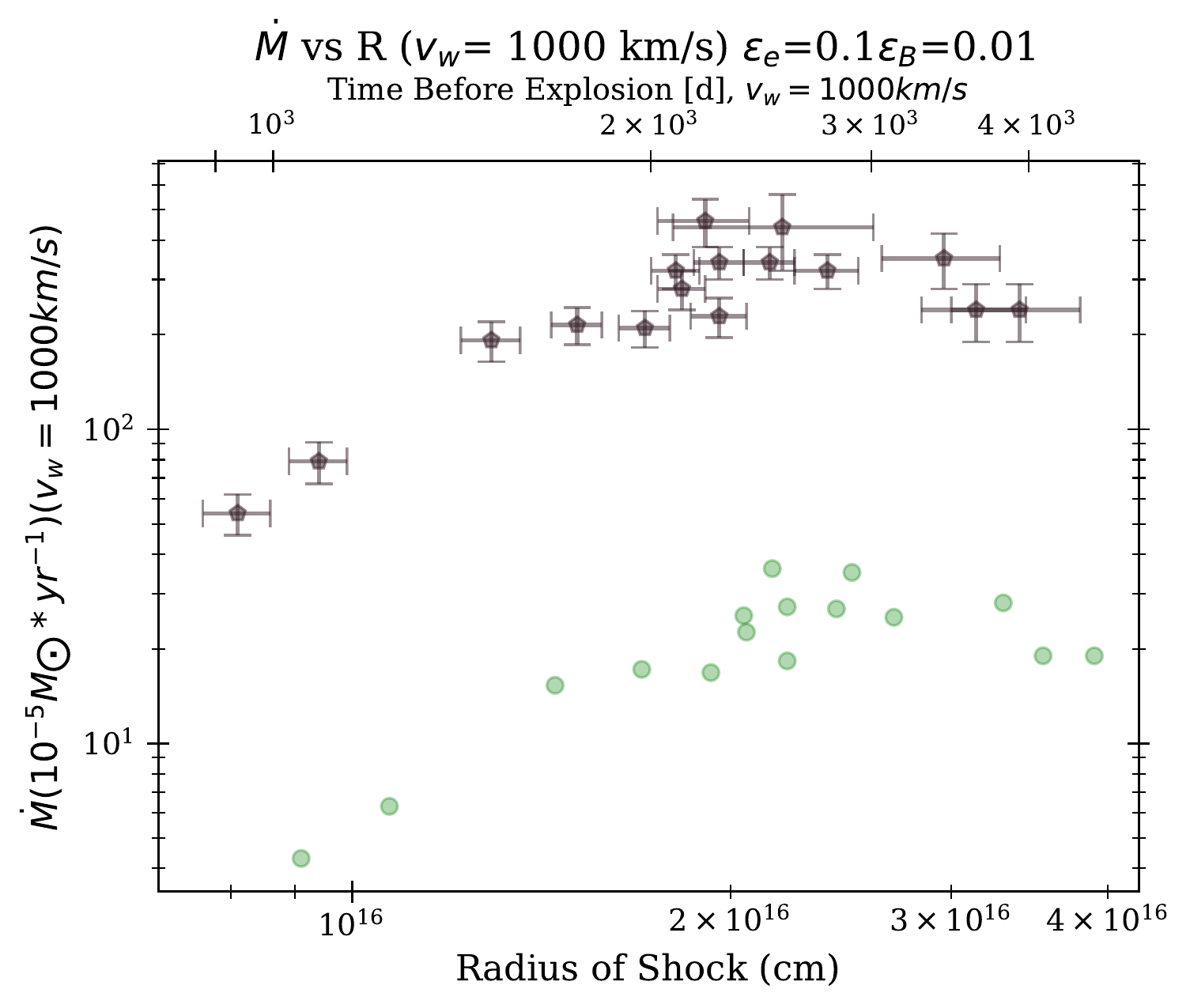}}\par 
{\includegraphics[width=.47\linewidth]{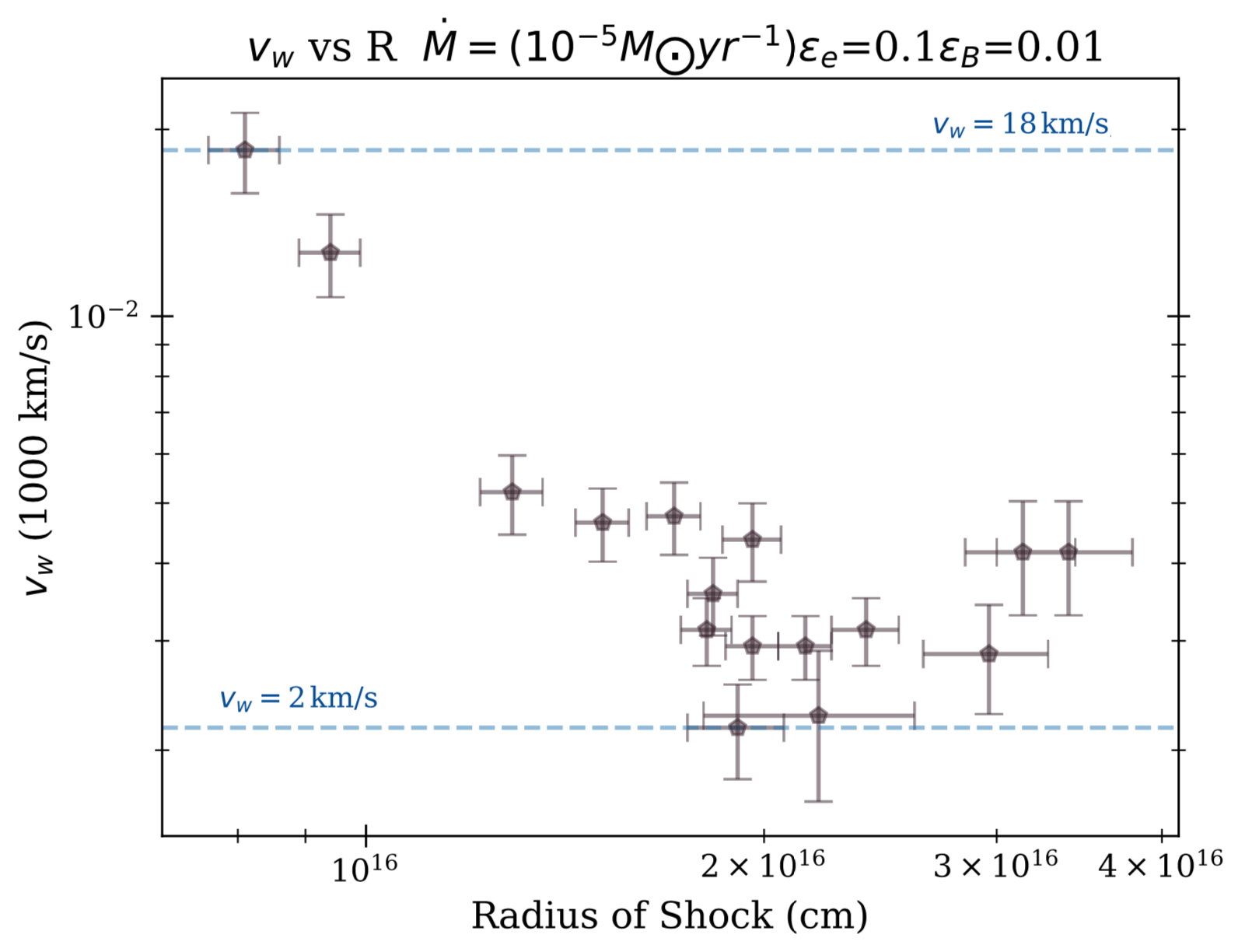}}
\caption{Physical parameters derived from $F_{\rm{brk}}$ and $\nu_{\rm{brk}}$ using our re-derived Equations \ref{eq:U} and \ref{eq:ML} for different choices of microphysical parameters. \textbf{Purple points} correspond to the favored broken power law model (\fitcase) with $\epsilon_B = 0.01, \epsilon_e =0.1$. \textbf{Green points} are in equipartition ($\epsilon_B = \epsilon_e =1/3$) and have the same uncertainties as the purple points.
\textbf{Top Panel, left:}  Blastwave energy U(t). 
\textbf{Top Panel, right:}  Mass loss assuming a wind velocity of $v_{w}=1000 \kms$ as a function of forward shock radius, which probes the CSM as it expands. The top abscissa denotes the time before explosion this material left the progenitor assuming $v_w = 1000\kms$.
\textbf{Bottom Panel:} Progenitor wind velocity assuming $\dot{M} = 10^{-5}$ \sm\ yr$^{-1}$ as a function of forward shock radius. \textbf{Dashed Lines:} The lowest and highest wind velocities in the dataset for this model and choice of microphysical parameters.}
\label{fig:mvu}
\end{figure*}

Figure \ref{fig:brv} presents the inferred evolution of the magnetic field with shock radius $B(R)$: 
we find evidence for a clear change in the $B$ evolution at $R_{br}\approx$ \breakradius cm, with a flat profile $B\propto R^{\text{\firstbvsrslope}}$ at $R<R_{br}$  evolving to $B\propto R^{\text{\secondbvsrslope}}$ at $R>R_{br}$.
As a reference, SN shocks expanding in wind-like media where $\epsilon_e$ and $\epsilon_B$ are not time-varying quantities are expected to show $B\propto R^{-1}$ \citep{chev98}.  
The location of this ``break radius''  is largely independent from the choice of microphysical parameters (Figure \ref{fig:brv}), due to the very mild dependence of $R$ on $\epsilon_e$ and $\epsilon_B$. 

We end this section by noting that in the context of SSA radio spectra for which the data provide constraints on only one break frequency $\nu_{\rm sa}$, the parameters $p$, $R$ and $B$ are the three physical properties that can be independently constrained from the data. Any other parameter, such as those discussed in the next section, are derived from combinations of $p$, $R$ and $B$. Furthermore, the parameters sensitively depend on assumptions on the shock microphysical parameters. The implications of different definitions of the microphysical parameters on calculations of $B$, as well as the effect of the BPL model chosen, are elucidated in Appendices \ref{app:ftest} and \ref{App:microphysics}.

\subsection{Inferred physical parameters of SN\,2004C: $U$, $\rho_{\rm{CSM}}$, $\dot M$, and $v_w$}
\label{SubSec:04CphysicsUrho}
In our calculations of the physical parameters that follow, we utilize the value of $q$ associated with the power-law slope of $R$ vs.\  $t$.
We use Equations \ref{eq:U}, \ref{eq:rho}, and \ref{eq:ML} to calculate $U(t)$, $\rho_{\rm{CSM}}(R)$ and $\dot M (R)$. 
We find that in equipartition the blastwave energy $U(t)$ increases with time from \lowestUequi ~erg at 41 days since explosion to \highestUequi ~erg at \dayshighestUequi ~since explosion (Figure \ref{fig:mvu}, green points).
This behavior is consistent with the progressive conversion of ejecta kinetic energy into shock internal energy, as the stratified ejecta decelerates into the environment. 
These values are lower limits on the true shock internal energy because of the likely deviation from equipartition.
However, as long as the shock microphysical parameters are constant with time, the ratio between the final and initial $U(t)$ is preserved, as is the temporal behavior of $U(t)$ or $\rho_{\rm{CSM}}(R)$.

\begin{figure*}[!ht]
    \centering
    \includegraphics[width=100mm]{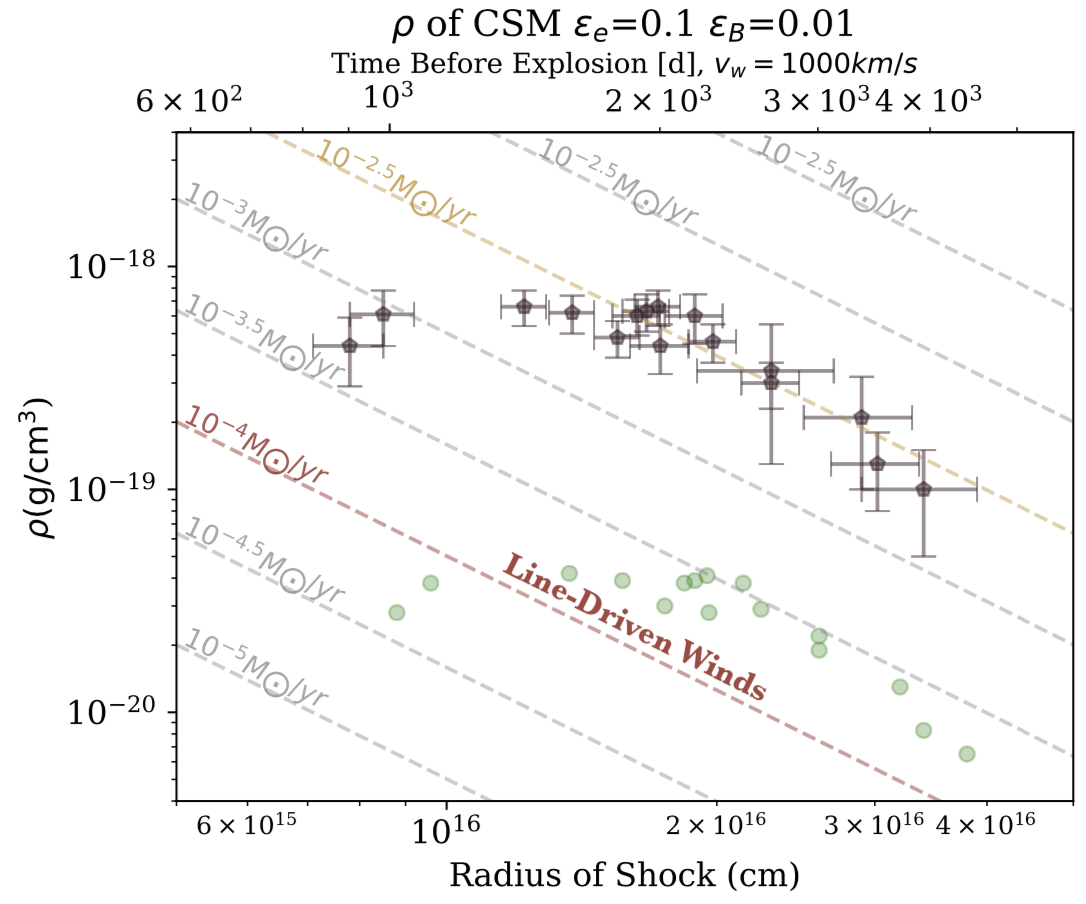}
    \caption{
    CSM density probed by the forward shock (upstream) as revealed by synchrotron emission at radio frequencies using our re-derived Equations \ref{eq:R} and \ref{eq:rho}. \textbf{Purple points} refer to values calculated using $\epsilon_e = 0.1$ and $\epsilon_B=0.01$; and \textbf{green points} refer to values calculated in equipartition, $\epsilon_e=\epsilon_B=1/3$; they will have the same uncertainties as the purple points. \textbf{Diagonal lines:} comparison to constant mass loss assuming $v_{w}=1000$ \kms. The mass loss-rate of \sn\ is much greater than can be explained by line-driven winds (\textbf{red dashed line,} $\dot{M}=10^{-4}$ \sm\ yr$^{-1}$) even at extreme luminosities $L=10^{6} \, L_{\odot}$. The \textbf{yellow dashed line} denoting $\dot{M} = 10^{-2.5}$ \sm\ yr$^{-1}$ is to assist the reader's estimate of the slope. \textbf{Top abscissa:} time before explosion material was deposited in the CSM, assuming $v_w=1000$~\kms\ and free expansion. }
    \label{fig:RhovsR}
\end{figure*}

With reference to Figure \ref{fig:RhovsR}, our analysis highlights the presence of an internal region with roughly constant density extending to $R_{\rm br}\approx $\breakradius\ cm, followed by a rapidly declining density profile with  $\rho_{\rm CSM}\propto R^{\text{\secondrhoslope}}$ at $R>R_{\rm br}$. 
For a nominal ejection velocity of the CSM material by the progenitor of $v_w=1000$ \kms, $R_{\rm br}$ corresponds to a look-back time of $\approx$ \timebreakMLRadius\ prior to stellar collapse.
Importantly, the $\rho_{\rm{CSM}}(R)$ profile that we derived from the radio observations violates the expectation from a single wind density profile, which predicts $\rho_{\rm{CSM}}\propto R^{-2}$. 
Our results point instead at the presence of a shell-like density structure $\sim 10^{16}$ cm from the explosion site of \sn\ not unlike other stripped-envelope SNe such as SN\,2004dk, ASASSN15no, SN\,2017dio, and SN\,2014C (\citealt{Margutti14C, benetti18, kuncara18, stroh21, bala21}, Brethauer, in prep.).
Interestingly, the phenomenology of mass lost to the environment in the last moments of evolution extends all the way to SLSNe (i.e., iPTF15esb,  iPTF16bad, SN\,2017ens), which presumably have different progenitors \citep{yan17, Chen18}.

Radio observations constrain the ratio $\rho_{\rm{CSM}}\propto \dot M/v_w$ but do not independently constrain $\dot M$ and/or $v_w$. 
It follows that the shell-like density profile around \sn\  can be the result of a time-varying mass-loss rate $\dot M(t)$ or a time-varying  progenitor wind $v_w(t)$, or both. 
Figure \ref{fig:mvu} (upper panel) shows the $\dot M$ profile as a function of radius and look-back time, if the progenitor had a roughly constant $v_w \sim 1000$  \kms.
Taken at face value, these results imply that in this case the very last phases of evolution of the progenitor at $t<$\timebreakMLRadius
~have been characterized by a rapidly decreasing $\dot M$ with time, with $\dot M$ reaching a peak value of $\dot M\approx$ \maxmassloss\ (for $\epsilon_e=0.1$ and $\epsilon_B = 0.01$).
We note that 
$\frac{\dot{M}}{v_w} \propto e_B^{\frac{8}{(13+2p)} - 1} e_e^{\frac{-8}{(13+2p)}}$ (Equation \ref{eq:ML}), so assuming equipartition versus the fiducial values of $\epsilon_e$ and $\epsilon_B$ can reduce the mass loss inferred by nearly a factor of $\sim$3 (Figure \ref{fig:mvu}). 
We report our results in Tables \ref{tab:phys_vals} and \ref{tab:phys_vals_equi}  in units of \constML\ and $v_w=$1000 \kms\ as these values are representative of WR stars in our galaxy ($\dot M=10^{-5.6} - 10^{-4.4}$ \sm\ yr$^{-1}$ and $v_w<6000$ \kms\, \citealt{crowther2007}).

For the complementary case of a constant $\dot M\approx 10^{-5}$ \sm\ yr$^{-1}$ (Figure \ref{fig:mvu}), the observed density profile would instead imply a significant change in the wind velocity in the last thousands of years preceding core-collapse, with a rapid increase of one of magnitude  in the final $\sim 3000$ years (from \lowwind\ to \lastwind\ for our assumed $\dot M$ unit value). 
These winds' velocities are clearly not consistent with a compact, single WR star.

In reality, neither of these two limiting cases might apply. The shell-like structure might be the result of a ``wind-wind interaction'', where faster, lighter winds interact with mass lost to the environment through slower winds of the preceding phase of stellar evolution, for example during the Red Supergiant (RSG) $\rightarrow$ WR transition, 
\citep{heger03, dwarkadas10, vanloon05,Mauron11}. 
We explore the ``wind-wind interaction'' scenario and additional explanations in the following section.

\section{Mass Loss Discussion}
\label{sec:massloss}

The major result of modeling our radio observations is a radial profile in $\rho_{\rm{CSM}}(R)$ that is robust against the choice of SED fit model (Figures \ref{fig:RhovsR} and \ref{fig:RhovsRall}). Most importantly, this profile reveals a structure in the CSM that is flat ($\rho \propto R^{\text{\firstrhoslope}}$) out to \breakradius\ cm, followed by an immediate, steep decline of $\rho \propto R^{\text{\secondrhoslope}}$. 

The maximum mass-loss rate that can be supported by line-driven winds depend on metallicity $Z$ and the luminosity $L$ of the progenitor star as follows \citep[e.g.,][]{smithowocki06,Gayley95}:
\begin{equation}
\label{Eq:Mdotmax}
    \dot{M}_{\rm{max, lines}} = 7 \times 10^{-3} Z L_6 M_{\sun}\rm{yr}^{-1},
\end{equation}
where $L_6 \equiv 10^6 \, L_{\odot}$.
For solar composition,  the hydrogen, helium, and metal fractions by mass are X$_\odot$ = 0.7381, Y$_\odot$ = 0.2485, Z$_\odot$ = 0.0134, respectively \citep{Asplund09}.
\host, the host galaxy of \sn, has approximately solar metallicity of about $12+[O/H] = 8.74$ dex
\citep{kelly2012}.\footnote{At the explosion site there is evidence for super-solar metallicity (Dittman, private communication).}  For the most luminous stellar progenitors with $L_6\sim 1$ at solar metallicity, 
Equation \ref{Eq:Mdotmax} implies 
$\dot{M}_{\rm{max, lines}}\approx 7 \times 10^{-5}$ \sm\ yr$^{-1}$, but \sn\ shows a maximum mass-loss rate of \maxmassloss (assuming $v_w =$ \WRwind, $\epsilon_B=0.01$, and $\epsilon_e=0.1$), two orders of magnitude greater than the limit of single line-driven winds. We conclude that a single fast wind ($v_w=1000$ \kms) from a compact progenitor cannot sustain the large mass-loss rates inferred for \sn.

\begin{figure*}
    \centering
    \includegraphics[width=120mm]{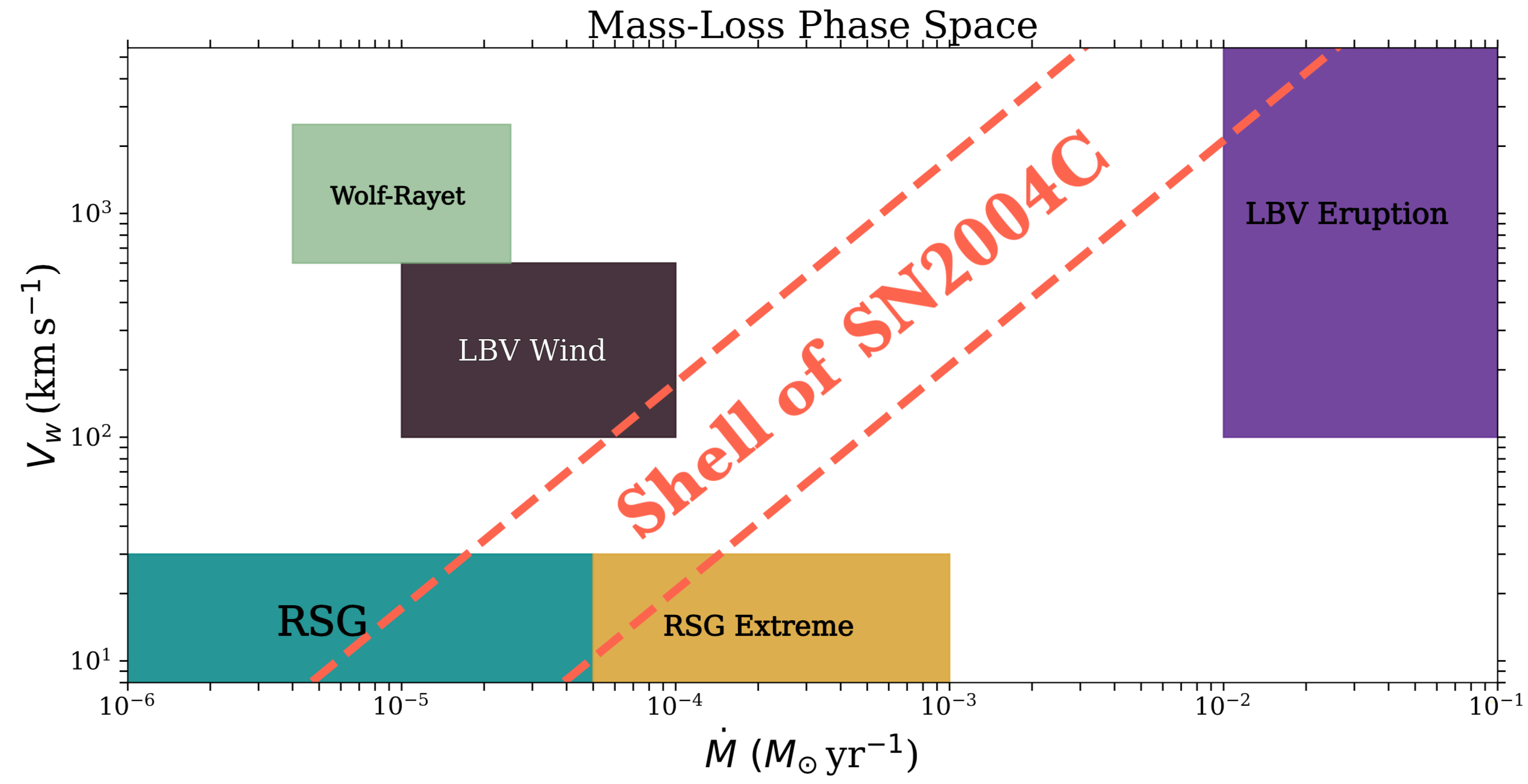}
    \caption{ 
    Wind velocity vs.\  mass-loss rate phase space (see \S\ref{sec:massloss} for full discussion). 
    \textbf{Blue Shaded Region:} Mass loss rates of main sequence stars and subgiants. Rates are taken from \citet{sea93}.
    \textbf{Light Green Shaded Region:} Mass loss rates of Wolf-Rayet stars from \citet{crowther2007}.
    \textbf{Brown Shaded Region:} Mass loss rates from LBV continuum winds \citep[and references therein]{smith2014, terreran19} 
    \textbf{Teal Shaded Region:} Wind speeds and mass loss rates of red supergiant stars \citep{sea93}.
    \textbf{Gold Shaded Region:} Extreme winds possible in red supergiants \citep{dejager88, marshall04, vanloon05}. 
    \textbf{Purple Shaded Region:} Observed mass losses of LBV eruptions \citep{smithowocki06}.
    \textbf{Orange Diagonal-Dashed Lines} demarcate the parameter space constrained by our observations of $\dot{M}/v_w$ in \sn. Note that although RSG winds fall within the realm of possibility, they have not been identified as SN Type IIb progenitors. 
    }
    \label{fig:singlewinds}
\end{figure*}

We now compare \sn\  to mass-loss rates observed in massive stars.  The densities measured for \sn\  overlap in the $\dot{M}$ vs.\ $v_w$ phase space with RSG and Luminous Blue Variable (LBV) progenitors, because RSGs have significantly slower winds and LBVs are not powered by line driven winds  (Figure \ref{fig:singlewinds}). However, both cases expect copious H in the ejecta, which was \emph{not} observed for \sn\ \citep{Shivvers17}. 
Assuming a WR wind speed of \WRwind\, the inner density profile of \sn\ corresponds to a mass loss rate of \MLinner\ at radius  $R=$\innerradius\ cm (with $\epsilon_B=0.01$ and $\epsilon_e=0.1$; Equation \ref{eq:ML}, Figure \ref{fig:mvu}).  WRs typically span a range of mass-loss rates $\dot M\approx 10^{-5.6}-10^{-4.4}$ \sm\ yr$^{-1}$, which is significantly lower than observed in \sn\ for the same wind velocity
\citep[compare WR box to \sn\ shell in Figure \ref{fig:singlewinds}]{crowther2007}.
Similarly, the region of \sn\  that overlaps with the parameter space partitioned to RSGs can be ruled out, as RSGs are H-rich and we did not see H in the spectra of \sn\ \citep{Shivvers17}. 
This comparison to known stellar mass-loss rates strengthens our previous conclusion that a single line-driven wind cannot produce the CSM material surrounding \sn; and instead we must invoke slower winds or multiple line-driven winds (Figure \ref{fig:singlewinds}).

As an alternative, continuum-driven winds can power larger mass-loss rates than line-driven winds due to their dependence on the electron scattering cross-section rather than the cross-sections of metals in the stellar atmosphere. 
In principle, this implies that continuum-driven winds are not as sensitive to metallicity. 
However, even if continuum-driven winds could manage to supply a constant mass-loss rate at the appropriate magnitude, we find the single-wind mechanism as a whole is in tension with the observed CSM density profile surrounding \sn.  
A single wind (line- or continuum-driven) with a constant mass-loss rate would create an environment density that scales with $\rho \propto R^{-2}$, rather than the measured $\rho \propto R^{\text{\firstrhoslope}}$ that transitions to $\rho \propto R^{\text{\secondrhoslope}}$ (Figure \ref{fig:RhovsR}). 

Instead, the density profile that we infer from radio observations requires us to consider more exotic forms of mass loss or wind-wind interaction scenarios. 
We investigate the possibility the CSM could have been sculpted by the interaction of multiple winds or time-varying winds in \S\ref{sec:windwind}. 
We explore the possibility that  this shell of material was ejected by a single event (i.e., a shell ejection scenario) in  \S\ref{sec:shellejection}.

\subsection{Wind-Wind Interaction and Time-Varying Continuum Winds} \label{sec:windwind}
Wind-wind interaction can occur in the circumstellar medium where a faster-moving wind encounters a slower-moving wind launched by the star at an earlier date. 
For reference, free expansion of a wind would produce a density profile $\rho \propto R^{-2}$, which is not too dissimilar from the outer shell density profile we observe in \sn, where $\rho \propto R^{\text{\secondrhoslope}}$ (Figure \ref{fig:RhovsR}).  
In the SN literature, wind-wind interaction models have been invoked to explain the ``anomalous'' density profiles found around SN\,1996cr \citep{dwarkadas10} and SN\,2001em \citep{chugai06}.  
\citet{chugai06} speculate that the shell of material around SN\,2001em resulted from a combination of late-stage mass transfer in a binary system followed by a RSG $\rightarrow$  WR transition. In this instance, both stages could could have polluted the environment and subsequently swept up the CSM as winds accelerated, then pushed the shell out to a radius of $\sim 10^{16}$ cm.
In the case of a single star, the transition to a compact WR phase (which is associated with a faster wind) is expected to take place $\approx 0.5 - 1$ Myr before explosion, much earlier than the approximate shell age of 1000 years in \sn\  \citep{dwarkadas10, Margutti14C}.
However, for some binaries, the envelope removal of the primary star by binary interaction can happen much closer to the epoch of core collapse, as it was proposed for SN\,2014C \citep{Margutti14C, danny14C}.
A similar connection between progenitor envelope and binary-induced mass loss was also proposed for Type IIb SNe by \citealt{maeda15}. 
It is possible that \sn\ resembles this scenario, requiring a binary system and a shorter WR phase than what single-star evolution allows. In this case, faster winds sculpt slower-moving winds in the CSM and the star consequently explodes as a stripped-envelope SN. 

Irrespective of the physical origin of the CSM shell, sustained mass loss in the final decades before explosion can significantly modify the stellar progenitor structure at the time of explosion. 
It is thus instructive to review our current knowledge of stellar progenitors of Type IIb SNe. 
Based on pre-explosion imaging with HST \citep{smartt15}, the prominence of the cooling-envelope phase in the optical light-curve \citep[e.g.,][]{Arcavi11}, and the dynamic of the fastest outflows launched by the explosions as constrained by radio observations \citep{Soderberg12}\footnote{Note however that the progenitor star of SN\,2011dh was likely extended, in spite of having launched an outflow with properties similar to those of the more compact Ib-Ic SN progenitors
\citep[see][and references therein.]{folatelli14}}, the stellar progenitors of Type IIb SNe show diversity in their structure, from extended yellow hypergiants like the recent SN\,2016gkg \citep{kilpatrick2017, tartaglia17, piro17,  orellana18, sravan18, kilpatrick2021} to more compact progenitors with $R_{\rm star}<50 \, R_{\sun}$, as in the case of Type IIb SN\,2008ax \citep[e.g.,][]{folatelli2015, yoon17, groh13, roming09, crockett08}.
Based on the current constraints on masses and radii of Type IIb progenitors (compiled from \citealt{sravan20, gilkis22} and references in both), the inferred escape velocities lie in the range $45 - 300$ \kms, which maps into typical stellar winds velocities of $135 - 900$ \kms\ 
(order of magnitude estimate). We note that the current sample of Type IIb SNe with detected progenitors is biased toward the more extended stars, and that compact stars that are more difficult to detect are associated with faster winds. In this sense, the estimates above should be viewed as a likely lower limits on the true wind velocity of \sn\  just before explosion. We thus conclude that for \sn, $v_w\gtrsim 100$ \kms\ at the time of the explosion, which implies large mass-loss rates $\dot M \gtrsim 10^{-4}$ \sm\ yr$^{-1}$.

\subsection{Shell Ejection Scenarios} \label{sec:shellejection}
Unlike steady winds, shell ejection scenarios mark an abrupt, sudden, and violent expulsion of mass from the progenitor, often triggered by an internal deposition of energy \citep{quataertandshiode, woosley15,  woosley17b,leung19a,leung20,wu21}.
In \sn, we observe a structure in the CSM between \innerradius\ cm and \outerradius\ cm. Assuming a constant wind speed of \WRwind, our model suggests this structure was produced over the course of \timeWRwindInner\ -- \timeWRwindOuter\ before core collapse (Figure \ref{fig:RhovsR}).
Therefore, we find evidence pointing to a shift in the behavior of the progenitor star in the final \percentlife\ of its lifetime.

Traces of violent, substantial outbursts depositing material $\sim 10^{16}$ cm from the star at the time of explosion have been reported across multiple SN types. 
Hydrogen-poor SNe Types Ib/c, IIb (such as \sn), and even SLSNe-I show evidence of CSM shells at this distance as we continue to monitor these events at late-time with radio observations  (\citealt{yan17, lunnan18, mauerhan18, benetti18, kuncara18, kundu19, gomez19, prentice20, jin21,  guti21, maeda21,   wynn19ehk, stroh21, Dong21}, Brethauer in prep., SN\,2019yvr Auchettl in prep., and the sampled population in \citealt{Griffin2021}).

An extreme case of continuum-driven winds is an LBV outburst \citep{humphreys17, Humphreys19}. 
These hot, blue stars quickly ionize stellar material, creating a large population of free electrons. In turn, the electrons serve as cross sections for radiative absorption and form metallicity-independent continuum-driven winds \citep{smith2014}.
Outbursts become progressively more extreme the closer a star gets to its Eddington limit, consequently providing a spectrum of mass-loss events and outbursts.
Such outbursts have been posited to source quasi-periodic mass loss observed in SNe at radio wavelengths (e.g. \citealt{kotak06}).
For example, $\eta$ Car's Great Eruption in the nineteenth century involved a loss of about $12-20$ \sm\ of stellar material. 
The majority of this material was lost over five years, but
a lower limit of about $\dot{M} = 0.5$ yr$^{-1}$
is estimated for the average over the $\approx$20 year duration of the eruption \citep{smithowocki06}.
Another example is the 1600 eruption of P-Cygni, wherein 0.1 \sm\ was lost, implying a rate of $\dot{M} = 10^{-2}$ \sm\ yr$^{-1}$ due to continuum winds \citep{owocki04, smithowocki06}.
These specific ejected mass and mass-loss rates are significantly larger than the $M\approx$ \shellsize\ CSM shell observed in \sn\ (Figures \ref{fig:RhovsR} and \ref{fig:singlewinds}).
While an LBV progenitor was suggested by quantitative modeling of the very early spectra of the Type IIb SN\,2013cu \citep{groh14} and the transitional Type Ibn/IIb SN\,2018gjx \citep{prentice20}, \sn, shows a clear difference from the historical LBV eruptions because its progenitor underwent core-collapse shortly after shell ejection. 
Furthermore, mass-loss rates of $>10^{-3}$ \sm\ are rare and have not been observed in living LBVs within the last few decades \citep{Davidson20}. In the following we explore shell-ejection mechanisms that are tied to the requirement of imminent core collapse.

In the case of ``Pulsational Pair-Instability'' SNe (PPISNe), stars with He cores $M >30$ \sm\ can undergo electron-positron pair production, removing pressure in their cores that would otherwise support the star against the crush of gravity. Consequential pulses rippling through the star can launch enormous amounts of mass at the stellar surface, or even the H envelope itself \citep[0.3 $-$ several \sm,][]{woosley17b}.
Because PPISNe exhibit multiple eruptions of unstable mass loss that mimic SNe, there is no way to know for certain if an explosion is a PPISN without observing a second eruption \citep{woosley17b}. Additionally, pre-SN core masses large enough for PPISNe require large Zero-Age Main Sequence (ZAMS) masses (e.g., 70--260 \sm), greatly favoring low-metallicity (sub-solar) environments \citep{woosley17b}. Because the host galaxy of \sn\  is about solar-metallicity
(with possibly a higher metallicity concentration at the site of the explosion (Dittman, private communication) we rule out the PPISN mechanism as an underlying explanation to our observations.

Another mechanism we consider is gravity waves, or wave heating. 
Proposed by \citealt{quataertandshiode}, hydrodynamic waves during late-stage nuclear burning can excite violent convection. At the onset of core carbon burning, neutrino cooling is unable to fully dissipate the generated energy and as a result, the extreme convection launches internal gravity waves with energies $\sim 10^7 \, L_{\odot}$ \citep{fullerandro18}. 
Barring large amounts of nonlinear wave breaking and damping throughout the body of the star (explored thoroughly in \citealt{wu21} and \citealt{fullerandro18}), the wave energy that survives to the surface can unbind mass, accelerate it above the escape velocity, and result in observable mass loss.

\citealt{fullerandro18} explore the wave heating mechanism in stripped and partially stripped progenitors at solar metallicity with 1D MESA models. 
Their models result in Type Ib/c and IIb SNe and place CSM material too nearby (within $6 \times 10^{14}$ cm) to apply to \sn. 
Similarly, \citet{wu21} conclude that stars of $M<30$ \sm\ could exhibit gravity-wave driven mass-loss outbursts in the years to decades before explosion, which is too late in the life of a star to be relevant for \sn.
Specifically, their H-poor models with $M_{\rm{ZAMS}} = 36-40$ \sm\ end their lives as $M \sim 15$ \sm\ WR stars. 
In this case, the energy generated by gravity waves was sufficient to eject $10^{-2}$ \sm, but the ejected mass only reached $r \sim 10^{14}$~cm before core collapse, significantly smaller than the inner radius of \innerradius\ we observe in \sn. 
Their model with the most distant CSM material ($r\sim10^{15}$cm) was an 11 $\sm$ yellow supergiant (YSG) progenitor, which lost 1 \sm\ over the 10 years prior to core collapse.
 1 $\sm$ is much larger than  the \shellsize\ observed in \sn\  between \innerradius\ and \outerradius\ cm. 
\citealt{Leung21b} also predict mass-loss rates of $10^{-5}-10^{-3}$ \sm$\rm{yr}^{-1}$\
out to $10^{12}-10^{14}$~cm in stripped progenitors $M_{\rm{ZAMS}}=(20-90)$ \sm\ and metallicites $Z=(0.002 - 0.02)$. 
These mass-loss rates are both too low, and the resulting CSM shells are too close-in to reproduce the situation of \sn. 
At the time of writing, current wave-driven models do not match the timescales and distances observed in \sn; and it is unclear if future simulations of wave-driven mass-loss could reproduce the measurements of \sn. 
We conclude that if related to wave-driven instabilities, the phenomenology that we observe in \sn\ requires this mechanism to be active earlier than nuclear burning stages.

As a result of pre-explosion imaging coupled with modeled interactions, binary star systems are one of the favored progenitor systems of SNe IIb \citep{sravan2019, yoon17, ryder18, fox14, zapartas2017}. 
The amount of stripping the primary star undergoes from the secondary companion strongly depends on the initial masses of the system, the initial separation of the two bodies, and their metallicities \citep{smitharnett14, yoon17}. 
It is believed that about $2/3$ of massive stars are members of binary systems that interact before stellar death, and both early and late Case B mass transfer (when the donor star is evolving into an RSG) can induce primary stars to become blue hypergiants, yellow hypergiants, and red supergiants -- all progenitors of SNe Type IIb  \citep{smitharnett14, yoon17, ryder18, fox14}.
We find that we cannot rule binary stripping and its possible effects on mass-loss winds throughout the lifetime of \sn, because binary interaction is capable of ejecting the entire hydrogen envelope of a star over a variety of timescales \citep{yoon17, ryder18, fox14}.

\section{conlcusions}
\label{sec:conclusion}

In this paper, we began with the synchrotron derivation of the radio emission from a SN shock by \chev\ and present here the expressions for the post-shock magnetic field $B$, forward shock wave radius $R$, shock internal energy $U$, circumstellar density $\rho_{\rm{CSM}}$, and associated mass-loss rate $\dot M$, generalized for any value of the power-law index of the electron density distribution, $p$.
As part of our generalized formalism, we reviewed the different definitions of the shock microphysical parameters $\epsilon_e$ and $\epsilon_B$ that are used in the radio SN literature and we show in detail how hidden assumptions can lead to systematically different inferences of the shock and environment parameters, even when the same microphysical parameters values are adopted. These discrepancies are summarized and quantified in Appendix \ref{App:microphysics}. We applied our generalized formalism of Equations \ref{eq:B} -- \ref{eq:ML} to a collection of 7.7 years of multi-frequency radio data of \sn\  and we showed how this formalism leads to inferences on the forward shock dynamics that are in agreement with our direct VLBI measurements.
Our modeling revealed the presence of a dense shell of CSM material containing a mass of \shellsize\ at a distance of about $10^{16}$~cm from the explosion. Our major results can be summarized as follows:

\begin{itemize}
    \item The shock microphysical parameters $\epsilon_e$ and $\epsilon_B$ have been defined in different ways in the radio SN literature. In this work we quantitatively show the impact of the most common definitions of these parameters on the inferred shock and environment parameters; and we present a parametrization that will enable a direct comparison of the inferences from different works in the literature. We emphasize that in the specific case of the mass-loss rates, different, typically hidden, definitions lead to $\dot M$ values that can differ up to a factor of $\sim5$ -- even for the same choice of value for the shock microphysical parameters. These are compared in Appendix \ref{App:microphysics}.
    \item Our radio modeling reveals a density profile with a flat structure in the CSM ($\rho \propto R^{\text{\firstrhoslope}}$) out to \breakradius\ cm, and an abrupt transition  to $\rho \propto R^{\text{\secondrhoslope}}$ at larger radii.  The most important result from the modeling of our radio observations is that a radial profile density $\rho_{\rm CSM }(R)$ in \sn\ that exists independently of the fitting model (Figures \ref{fig:RhovsR} and \ref{fig:RhovsRall}). 
    \item If the \shellsize\ of CSM material around \sn\ was created by a shell ejection event, the implied ejection epoch is $128-2.5$ years prior to explosion for an ejection velocities between $20-1000\,$\kms. 
    \item Of the mass-loss mechanisms we discuss in \S\ref{sec:massloss}, we assert a single, line-driven wind with constant speed and mass-loss rate \textit{could not} have sculpted the structure observed in the CSM of \sn\ (Figure \ref{fig:singlewinds}). Instead, multiple line-driven winds, a single time-varying wind, or a time-varying $\dot{M}$ could be responsible. We cannot rule out binary interaction as the underlying cause. While the extreme limits of gravity waves/wave heating and LBV mass-loss rates marginally overlap with the uncertainties of our calculated rates, current models predict material either too close to the star or released too slowly.
\end{itemize}

The connection between the shell characteristics, mass-loss mechanism, and timing of mass-loss events events highly suggests a fundamental aspect of stellar evolution in evolved massive stars at work.
Other SNe across different explosion types exhibit similarly substantial overdensities in their environments at $r\sim10^{16}$ cm from their explosion site \citep{Margutti14C, Griffin2021, benetti18, yan17, Leung21a, wynn19ehk, kuncara18, Chen18, wynn20tlf, stroh21},
possibly suggesting a shared mechanism.
By observing changes in radio emission from the SN shock wave to probe the CSM density, we gather observational evidence that a rich and complex mass-loss history exists in stars, implying episodic mass loss 
in their final moments while they are on the verge of collapse. 
Consequently, the inability to conclusively point to any particular mechanism as the underlying cause highlights the need to model the final moments of late-stage core burning in H-poor progenitors.

\acknowledgments
L.D. heartily thanks Avni Coda for her contribution to all data visualization and gratefully acknowledges the help of \url{tablesgenerator.com} for importing .csv files as \LaTeX\ tables with minimal headache. 
L.D. is grateful for the partial financial support of the IDEAS Fellowship, a research traineeship program funded by the National Science Foundation under grant DGE-1450006.
The National Radio Astronomy Observatory is a facility of the National Science Foundation operated under cooperative agreement by Associated Universities, Inc.
The Margutti team at UC Berkeley and Northwestern is supported in part by the National Science Foundation under Grant No. AST-1909796 and AST-1944985, by the Heising-Simons Foundation under grant \# 2018-0911. 
R.M. is a CIFAR Azrieli Global Scholar in the Gravity \& the Extreme Universe Program 2019, and a Sloan Fellow in Physics, 2019. 
W.J.-G. is supported by the National Science Foundation Graduate Research Fellowship Program under Grant No.~DGE-1842165. 
W.J.-G. acknowledges support through NASA grants in support of {\it Hubble Space Telescope} programs GO-16075 and 16500.
D.~M.\ acknowledges NSF support from grants PHY-1914448 and AST-2037297. 
Parts of this research were supported by the Australian Research Council Centre of Excellence for All Sky Astrophysics in 3 Dimensions (ASTRO 3D), through project number CE170100013.  Research at York University was supported by the Natural Sciences and Engineering Research Council of Canada.

\software{lmfit \citep{lmfit}, SNID \citep{SNID}, ParselTongue \citep{Kettenis06}}


\appendix

\section{Synchrotron Constants}
\label{App:constants}
We report the constants found in \citealt{P1970}, which are used throughout Section \ref{sec:derivation}. 
In the following, $p$ is the power-law evolution of the accelerated electrons distributed in electron Lorentz factor $\gamma_e$, with minimum Lorentz factor $\gamma_{\rm{m}}$: $dN(\gamma_e)/d\gamma_e = K_0 \gamma_e^{-p}$ for $\gamma_e \ge \gamma_{\rm m}$. $N$ is the number of electrons per unit volume and  $K_0$ is the normalization. Other constants in the following equations include the charge of the electron $e$, the mass of an electron $m_e$, the speed of light $c$, and the Gamma function $\Gamma(p)$. All constants are in c.g.s.\ units.

\begin{equation}
    c_1 = 6.27 \times10^{18}
\end{equation}

\begin{equation}
    c_5= \frac{\sqrt3}{16 \pi} \frac{e^3}{m_ec^2}\frac{p+\frac{7}{3}}{(p+1)}\Gamma \Big(\frac{3p-1}{12}\Big)\Gamma \Big( \frac{3p+7}{12} \Big)
\end{equation}

\begin{equation}
    c_6 = \sqrt3 \frac{\pi}{72}e m_e^5c^{10}(p+\frac{10}{3})\Gamma \Big(\frac{3p+2}{12} \Big)\Gamma \Big(\frac{3p+10}{12}\Big)
\end{equation}

\newpage
\section{Model Comparison}
\label{app:ftest}

The following Table \ref{tab:models} is a statistical comparison of the performance of different SED models.
We first employ the F-Test, which begins by calculating the F-statistic, generally expressed as:

\begin{dmath}
F= \frac{(\chi_1^2/dof_1)}{(\chi_2^2/dof_2)}.
\end{dmath}

Model 1 has a larger number of degrees of freedom (d.o.f.) and fewer free parameters ($M_{\rm params}$). Our Null Hypothesis ($H_0$) is that Model 1 is the preferred fit, following the argument of Occam's Razor. We reject $H_0$ if the $p$-value is $p<5\%$.

Our radio data set of \sn\ has a number of data points $N_{\rm data} = 94$.
The model names are abbreviated with ``BPL'', referring to their broken power law shape (Equation \ref{eq:bpl}) and they represent a joint fit of 28 radio SEDs which have 56 free parameters (i.e. $\nu_{\rm{pk}}, F_{\rm{pk}}$, one for each SED) with an additional 1 to 3 shared parameters between them (slope $\alpha_1$, slope $\alpha_2$, and smoothing parameter $s$). A full discussion of fit parameters and methods is provided in \S\ref{sec:fit} and \ref{sec:model}. 
For the comparison of BPL2 to BPL3, the number of degrees of freedom is the same, and we use the $\chi^2$ statistic. 
We identify BPL1 (\fitcase) as the favored model by these statistics.  

In the Table, Model 1 refers to BPL1, Model 2 refers to BPL2 etc. As an example of how to read the table, for BPL2 (which is Model 2), the column ``$\rm{F-Stat\,1}$'' reports the value of the statistics with respect to Model 1, while the column ``$\rm{p-value\,1}$'' reports the corresponding $p$-value.

\begin{deluxetable}{|l|l|l|l|l|l|l|l|l|l|l|l|l|}[h!]
\label{tab:models}
\tablecaption{SED models performance quantified with the F-test and $\chi^2$. ``d.o.f.'' indicates the number of degrees of freedom in the model. }
\tablecolumns{12}
\tablewidth{\textwidth}
\tablehead{
\thead{Model} &
\thead{$\chi^2$} &
\thead{$M_{\rm{params}}$} &
\thead{d.o.f.} &
\thead{F-Stat 1} &
\thead{$p-$value 1} &
\thead{F-Stat 2} &
\thead{$p-$value 2} &
\thead{F-Stat 3 } &
\thead{ $p-$value 3 } &
\thead{F-Stat 4} &
\thead{$p-$value 4} 
}
\startdata
\makecell{$\alpha_2 = 5/2$ \\ and $s=-1$ \\ (``BPL1'')} & 70.82 & 45 & 49 & $H_0$ & $H_0$ & - & - & - & - & - & -  \\ \hline
\makecell{$s=-1$ \\ (``BPL2")} & 61.89 & 46 & 48 & 1.12 & 0.34  & $H_0$ & $H_0$ & - & - & - & -  \\ \hline
\makecell{$\alpha_2 = 5/2$ \\ (``BPL3'')} & 64.19 & 46 & 48 & 1.08 & 0.39 & -- & -- & $H_0$ & $H_0$ & - & -  \\ \hline
\makecell{both slopes and \\ $s$ vary \\ (``BPL4'')} & 56.74 & 47 & 47 & 1.20 & 0.27 & 1.11 & 0.36 & 1.16 & 0.31 & $H_0$ & $H_0$ 
\enddata
\end{deluxetable}

\begin{figure}[ht]
\begin{tabular}{cc}
    \includegraphics[width=61mm]{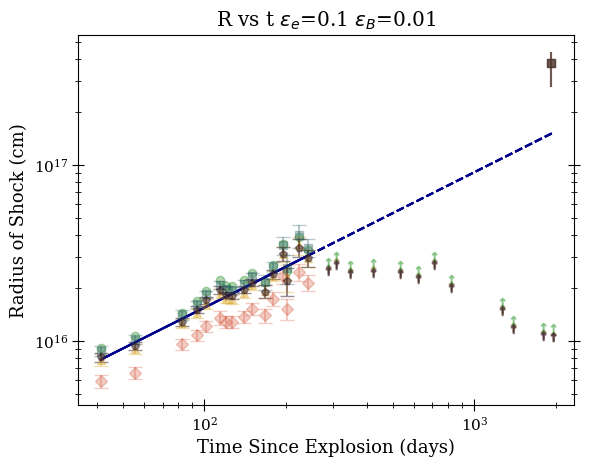} &
    \includegraphics[width=65mm]{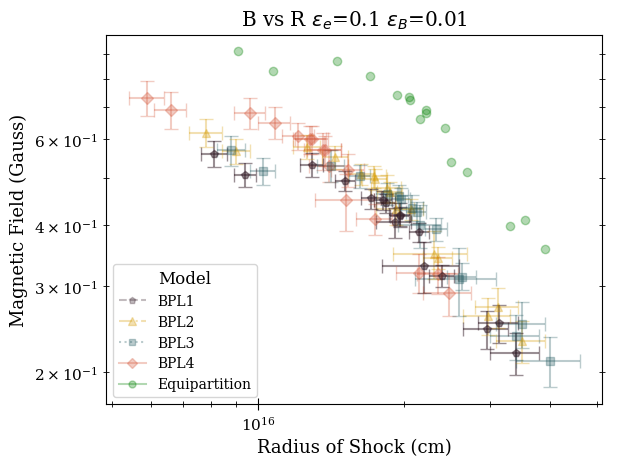} \\
    \includegraphics[width=61mm]{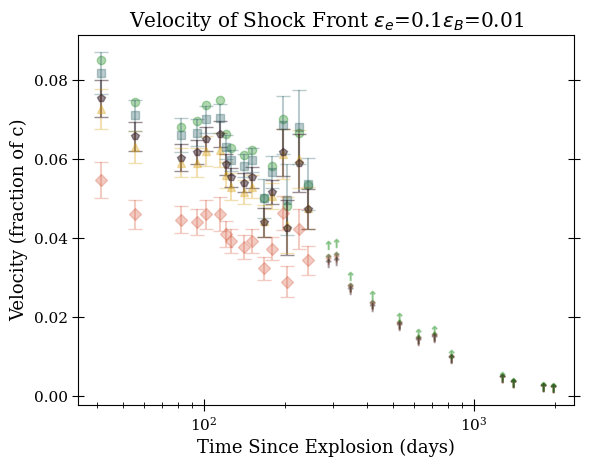} &
        \includegraphics[width=65mm]{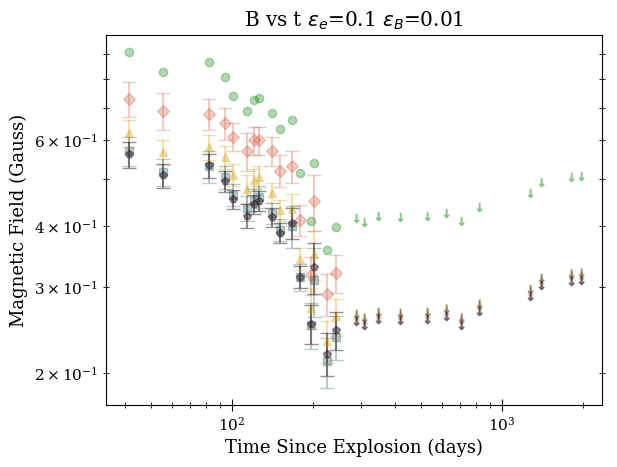}\\
\end{tabular}
\caption{The same as Figure \ref{fig:brv}, but with all models compared in Section \ref{sec:fit} and Appendix \ref{app:ftest}. Note that the normalization differs between models, but the overall evolution remains the same.  \textbf{Purple pentagons:} Favored model where smoothing parameter is fixed at $s=-1$ and the slope $\alpha_2=5/2$ (``BPL1''). \textbf{Yellow triangles:} only smoothing parameter is fixed $s=-1$ (``BPL2''). \textbf{Blue squares:} only slope is fixed $\alpha_2 = 5/2$ (``BPL3''). \textbf{Pink diamonds:} All slopes and $s$ are free to vary (``BPL 4''). \textbf{Green points:} BPL1 in equipartition ($\epsilon_B= \epsilon_e = 1/3$). \textbf{Top, left panel; blue line:} Power-law fit $R\propto t^{q}$, where $q=\text{\q}$ and appears in our generalized formulae as the variable $q$.}
\label{fig:all_models_brv}
\end{figure}

\begin{figure}[ht]\centering
{\includegraphics[width=.4\linewidth]{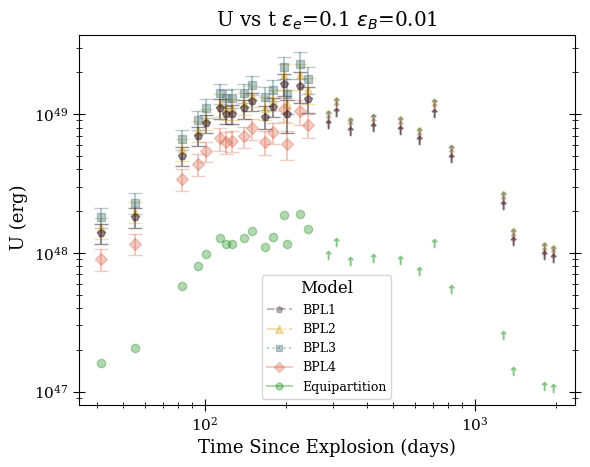}}\hfill
{\includegraphics[width=.4\linewidth]{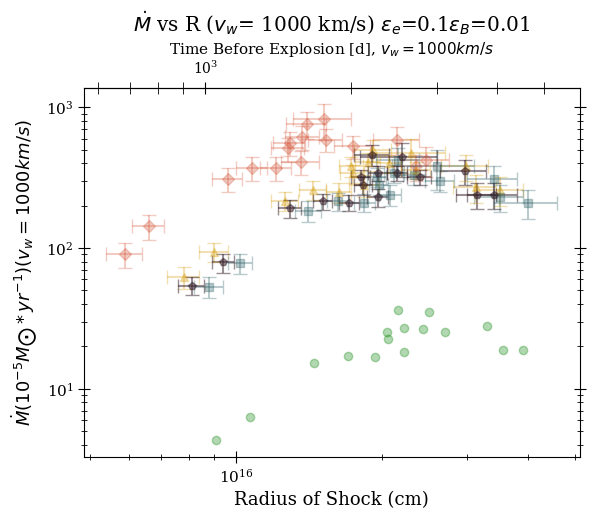}}\par 
{\includegraphics[width=.4\linewidth]{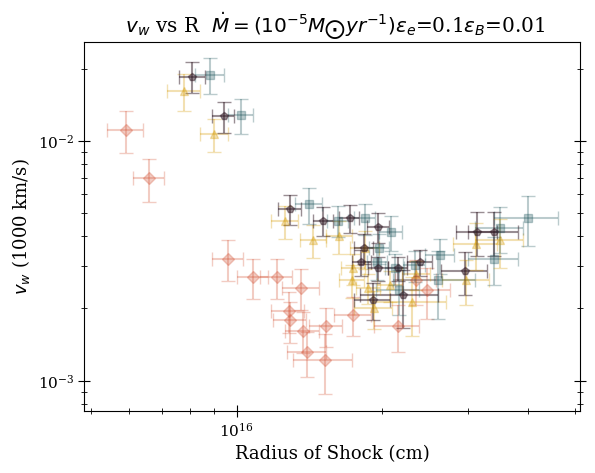}}
\caption{The same as Figure \ref{fig:mvu}, but with all models compared in Section \ref{sec:fit} and Appendix \ref{app:ftest}. Note that the normalization differs between models, but the overall shape of the evolution remains the same. \textbf{Purple pentagons:} Favored model where smoothing parameter is fixed at $s=-1$ and the slope $\alpha_2=5/2$ (``BPL1''). \textbf{Yellow triangles:} only smoothing parameter is fixed $s=-1$ (``BPL2''). \textbf{Blue squares:} only slope is fixed $\alpha_2 = 5/2$ (``BPL3''). \textbf{Pink diamonds:} All slopes and $s$ are free to vary (``BPL 4''). \textbf{Green points:} BPL1 in equipartition ($\epsilon_B= \epsilon_e = 1/3$).  }
\label{fig:all_models_mvu}
\end{figure}

\begin{figure*}[ht]
    \centering
    \includegraphics[width=100mm]{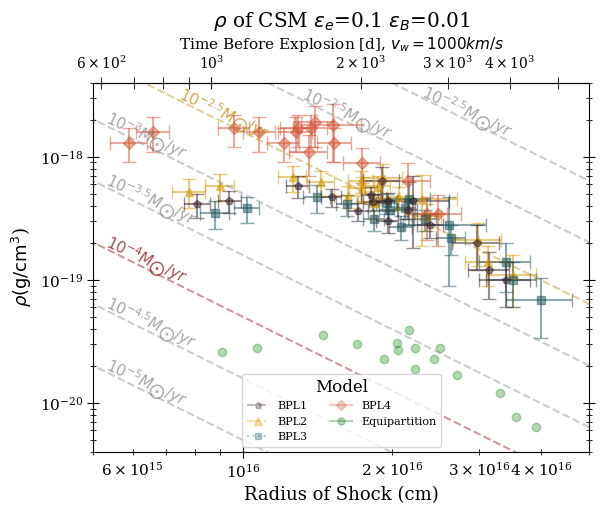}
    \caption{The same as Figure \ref{fig:RhovsR}, but with all models compared in Section \ref{sec:fit} and Appendix \ref{app:ftest}. Note that the normalization differs between models, but the overall shape of the evolution remains the same. \textbf{Purple pentagons:} Favored model where smoothing parameter is fixed at $s=-1$ and the slope $\alpha_2=5/2$ (``BPL1''). \textbf{Yellow triangles:} only smoothing parameter is fixed $s=-1$ (``BPL2''). \textbf{Blue squares:} only slope is fixed $\alpha_2 = 5/2$ (``BPL3''). \textbf{Pink diamonds:} All slopes and $s$ are free to vary (``BPL 4''). \textbf{Green points:} BPL1 in equipartition ($\epsilon_B= \epsilon_e = 1/3$).  }
    \label{fig:RhovsRall}
\end{figure*}

We also present Figures \ref{fig:all_models_brv} and \ref{fig:all_models_mvu} as extensions of Figures \ref{fig:brv} and \ref{fig:mvu}, respectively; and we provide a version of Figure \ref{fig:RhovsR} in Figure \ref{fig:RhovsRall}. These figures contain all four models as well as the equipartition case of BPL1 for ease of comparison. We note that the choice in model can cause a spread amongst values of physical parameters, but the overall evolution of the inferred parameters remains the same. Additionally, we highlight that the largest variation is associated with the choice of micro-physical parameter values $\epsilon$. 

\newpage

\newpage

\newpage

\newpage
\newpage
\newpage
\section{Definitions of the shock microphysical parameter $\epsilon_B$ (and $\epsilon_e$) }
\label{App:microphysics}
In the radio SN literature the parameter $\epsilon_B$ (and $\epsilon_e$)\footnote{The definitions of $\epsilon_e$ can be easily obtained by replacing $E_B=B^2/8\pi$ with $E_e$, and $\epsilon_B$ with $\epsilon_e$ in Equations \ref{Eq:epsilonBchevalier}-\ref{Eq:epsilonBP2}.
} has been defined in at least four different ways:
\begin{equation}
    \label{Eq:epsilonBchevalier}
  \frac{B^2}{8\pi} \equiv \epsilon_B \rho_{\rm CSM}v^2 
\end{equation}

\begin{equation}
    \label{Eq:epsilonBMatsuoka}
   \frac{B^2}{8\pi} \equiv \epsilon_B \rho_{\rm CSM,sh}v^2 = \frac{\Gamma+1}{\Gamma-1} \epsilon_B \rho_{\rm CSM}v^2 \setg 4 \times \epsilon_B \rho_{\rm CSM}v^2 
\end{equation}

\begin{equation}
    \label{Eq:epsilonBUthermal}
   \frac{B^2}{8\pi} \equiv \epsilon_B E_{th} =
   \epsilon_B
   \frac{1}{2}
   \rho_{\rm{CSM,sh}}
   \bigg[
   \bigg(1-\frac{\Gamma-1}{\Gamma+1}\bigg)
   v 
   \bigg]^2  \setg
   \frac{9}{8}\times \epsilon_B \rho_{\rm{CSM}}v^2 
\end{equation}

\begin{equation}
    \label{Eq:epsilonBP2}
   \frac{B^2}{8\pi} \equiv \epsilon_B P_2 = \frac{2}{\Gamma+1} \epsilon_B \rho_{\rm{CSM}}v^2
   \setg \frac{3}{4}\times \epsilon_B \rho_{\rm{CSM}}v^2 
\end{equation}

\noindent
where $B$ is the magnetic field in the shocked CSM, $v$ is the FS velocity, $\rho_{\rm CSM }$ ($\rho_{\rm CSM,sh}$) is the pre-shock (post-shock) CSM density, $\Gamma$ is the adiabatic index, $E_{\rm th}$ is the post-shock thermal energy density, and $P_2$ is the post-shock gas pressure. 
In other words, $\epsilon_B$ is defined as a fraction of different energy densities that in this paper we parametrize as $ \frac{2}{1+i}\rho_{\rm CSM}v^2 $. The corresponding energy is $U$ (Equation \ref{eq:U}).

In the limit of a strong shock scenario the traditional Rankine–Hugoniot jump conditions apply: $\rho_{\rm CSM,sh}= \frac{\Gamma+1}{\Gamma-1} \times \rho_{\rm CSM}$.
For an ideal  monatomic gas, $\Gamma=5/3$, which leads to $\rho_{\rm CSM,sh}= 4 \times \rho_{\rm CSM}$,  $E_{\rm th}= \frac{9}{8} \rho_{\rm CSM}v^2$, and $E_{\rm th}=\frac{P_2}{\Gamma-1}=\frac{3}{2}P_2$.  In this paper we use $\frac{B^2}{8\pi}=\frac{2}{i+1} \epsilon_B \rho_{\rm  CSM}v^2$ as a generic definition, where the parameter $i$ 
provides flexibility between the different definitions above. When one chooses  definition \ref{Eq:epsilonBP2}, $i$ in our equations becomes conveniently equivalent to the adiabatic index, $\Gamma$.
\citet{chev98,chevfrans17} and related literature adopt the definition of Equation \ref{Eq:epsilonBchevalier}.  Definitions \ref{Eq:epsilonBMatsuoka}, \ref{Eq:epsilonBUthermal}, \ref{Eq:epsilonBP2} have been employed, for example, in \citet{matsuoka2020}, \citet{petropoulou16}, and \citet{Ho19}, respectively. While we present the final inferences on the physical parameters of SN\,2004C following \chev,  we leave the $i$ explicit in our equations to easily visualize the impact of the different definitions on the estimates of the physical parameters.

The obvious consequence of these different definitions of the microphysical parameters is that the inferred values of the physical parameters of the system (e.g., $B$, $R$, $\rho_{\rm{CSM}}$) are not directly comparable among different works, even when the same $\epsilon_B$ (and $\epsilon_e$) values are assumed or derived.
Derived values of $\dot{M}$ can differ by as much as a factor of $\sim 5$ in this way. 

In addition, there are two approaches used in the literature for the FS velocity $v\equiv dR/dt$. The first approach utilizes an averaged FS shock velocity since explosion: 

\begin{equation}
    \label{Eq:vshockwave}
   v\approx \frac{R}{t} \equiv v_{\rm avg}
\end{equation}
A second approach assumes a (local or global) power-law evolution of the FS radius with time $R(t)\propto t^q$:
\begin{equation}
    \label{Eq:vshockPL}
   v\equiv dR/dt=q\frac{R}{t}
\end{equation}

At early times the typical non-relativistic SN FS is slowly decelerating and the assumption of a linear evolution of the shock radius with time implicit in Equation \ref{Eq:vshockwave} is approximately correct. 
For SNe in the interaction phase expanding in a power-law density medium for which the self-similar solutions of \citet{chev82} apply, Equation \ref{Eq:vshockPL} provides the exact solution. 
Following \citet{chev82}, $q=\frac{n-3}{n-s}$, where $\rho_{\rm CSM}\propto r^{-s}$ and the SN ejecta outer density profile  is $\rho_{\rm{ej}}\propto r^{-n}$. 
For a wind density profile $s=2$, and a compact massive star (i.e., $n\approx 10$) such as typically assumed for the progenitors of H-stripped SNe, $q=\frac{7}{8}$.  
Definitions of Equations \ref{Eq:epsilonBchevalier}-\ref{Eq:epsilonBP2} have been combined with either Equation \ref{Eq:vshockwave} or \ref{Eq:vshockPL} to solve for the parameter of interest $\rho_{\rm CSM}$ (and, by extension, $n_e$ and $\dot{M}$). For example, \citet{chevandfrans} adopt Equation \ref{Eq:vshockPL} for $q=\frac{7}{8}$, while \citet{Ho19} uses the definition of Equation \ref{Eq:epsilonBP2} with the averaged FS velocity $v_{\rm avg}$ of Equation \ref{Eq:vshockwave}. 
In this paper we assume a local power-law evolution of the FS radius with time without assuming values for $n$ or $s$.  
Instead, we estimate $q$ from a local fit of the FS $R(t)$ inferred from radio data following  Equation \ref{eq:R}.
In our equations we leave $q$ as a variable, and note that Equation  \ref{Eq:vshockPL} reduces to Equation \ref{Eq:vshockwave} for the special case $q=1$.

\newpage
\section{Radio Data and Calculations } \label{App:dataandcalc}

\begin{figure}[ht]
\centering
   \includegraphics[width=120mm]{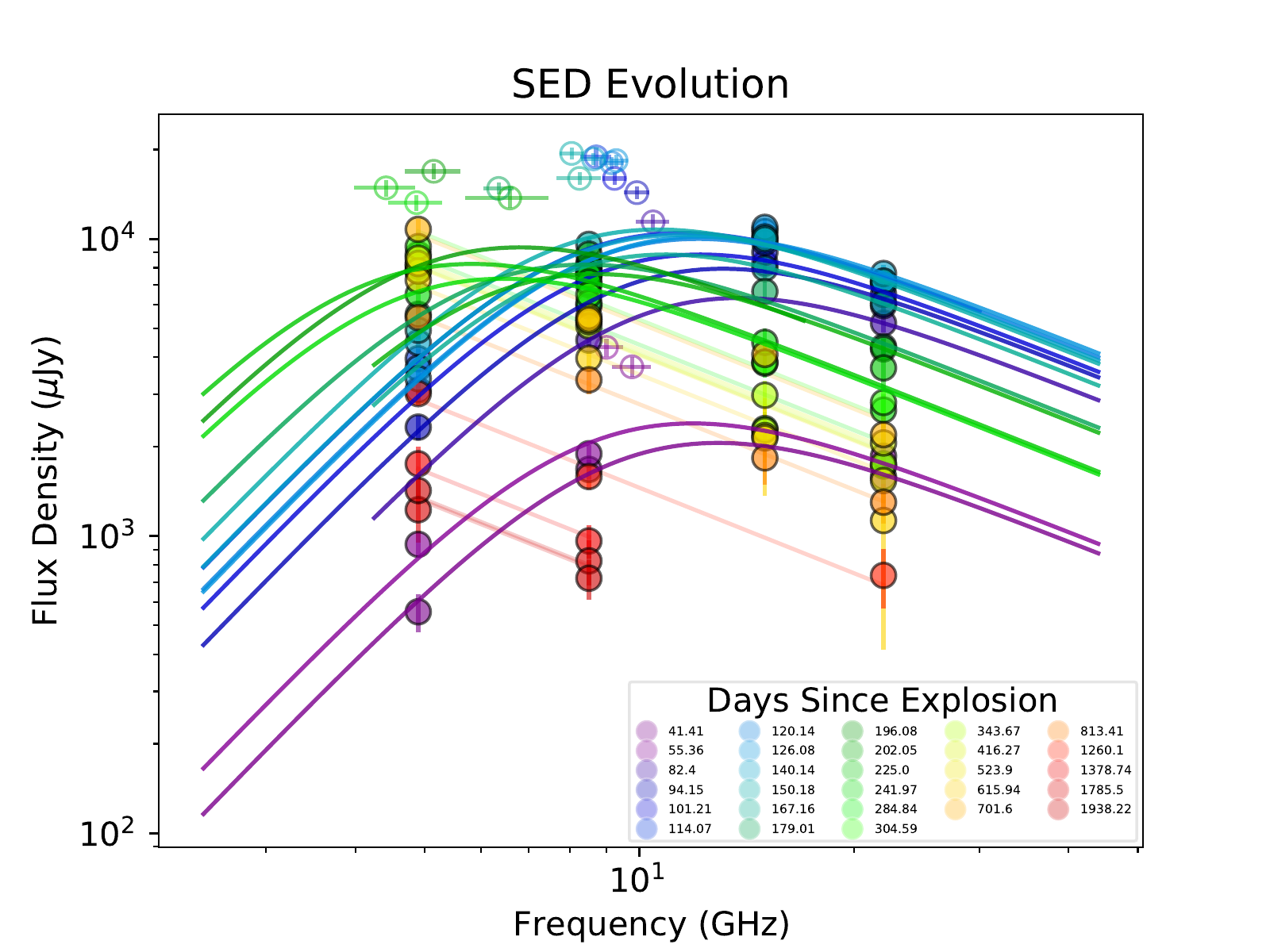}  
\caption{The full spectral energy distribution (SED) evolution of \sn, with data grouped in SED following the criterion of time range of acquisition $\delta t / t < 0.025$. \textbf{Solid points:} radio data of \sn\ (provided in Table \ref{tab:data}). \textbf{Solid lines:} our favored model BPL1, a broken power law with smoothing parameter $s$ fixed to $-1$ and optically thick slope $\alpha_2$ fixed to $5/2$. \textbf{Hollow points:} the asymptotic intercept of the thick and thin slopes of the broken power law, identified as $F_{\rm{br}}$ and $\nu_{\rm{br}}$ in Table \ref{tab:fit} and used in our calculations of the physical parameters. At later times, the SED peak migrates to frequencies too low to capture both sides of the SED. In this case, a single power law is fit to the SED and the peak becomes a flux density (frequency) lower limit (upper limit) propagated through the physical parameters.
}
\label{fig:SED}
\end{figure}

We provide the full radio dataset of \sn\  in Table \ref{tab:data}.
Similarly, we provide in Tables \ref{tab:phys_vals} and \ref{tab:phys_vals_equi} the physical parameters calculated using the best fit parameters inferred from the modeling of the 28 SEDs modeled with BPL1 (Table \ref{tab:fit}) and Equations \ref{eq:B} -- \ref{eq:ML}. These results are illustrated in Figures \ref{fig:brv}, \ref{fig:mvu}, and \ref{fig:RhovsR}. Figure \ref{fig:SED} shows the best fit model BPL1 applied to the full data set.

\input{data/physical_values}

\end{document}

%% file: sn2004c_v2.tex
\subsection{Explosion date and SN typing }
\label{sec:optical}
SN\,2004C was discovered January 12.5, 2004 UT in near-infrared K'-band (2.1$\mu$m) imaging with relatively poor constraints on the time of explosion \citep{Dudley04}. The most recent prior observation of that region was taken May 18, 2003 and published photometry are insufficient to robustly constrain the time of maximum light and potentially extrapolate an explosion date. The only known attempt was made by \citet{Meikle04}, who estimated that SN\,2004C was 3 weeks past maximum light as of January 16, 2004 by comparing its $g^{\prime}-r^{\prime}$ and $r^{\prime}-i^{\prime}$ colors to those of Type Ic supernovae. That estimate, however, is problematic given that years later \citet{Shivvers17} reported that the spectral classification of SN\,2004C was in fact Type IIb, and not Type Ic as originally reported \citep{Matheson04}. 

To better estimate the date of explosion, we supplemented these reports and the updated spectroscopic classification with our own investigation of archival optical spectra of SN\,2004C retrieved from WISeREP \citep{WISEREP}. Spectra spanned January 15, 2004 to March 17, and were inspected for their likeness to other Type IIb spectra tagged with time since explosion and/or maximum light. Our comparisons were aided with use of \texttt{SNID} \citep{SNID}; and we heavily weighted a high quality Lick 3m spectrum of SN\,2004C obtained January 17, 2004 and originally published in \citet{Shivvers17} in our manual comparisons. Among all Type IIb supernova spectra inspected, those of SN\,1993J and SN\,2000H displayed the closest similarity in spectral evolution with the limited available epochs of SN\,2004C. Comparisons converged on the time of maximum light of SN\,2004C to be around January 2, 2004.  Given an average Type IIb V-band rise-to-peak time of approximately 18 days from explosion (see, e.g., \citealt{Milisavljevic13}), we estimate the explosion date to be December 15, 2003 with an uncertainty of $\sim$\,2 weeks dominated by ambiguity between fits with other epochs. In the following we adopt MJD 52988 (corresponding to December 15, 2003) as the explosion date. This uncertainty has no impact on our major conclusions.

%% file: data/physical_values.tex
\begin{deluxetable}{|l|l|l|l|l|c|}
\tablecaption{Radio observations of \sn. Flux densities are in $\mu$Jy. The measurement on 54919.0 with the VLBA was taken at 8.4 GHz (Section \ref{sec:obs}). Reported are only the rms uncertainties. }
\label{tab:data}
\tablecolumns{6}
\tablehead{
\thead{Date (MJD)} &
\thead{$F_{\nu}$ 4.9 GHz} & 
\thead{$F_{\nu}$ 8.5 GHz}& 
\thead{$F_{\nu}$ 15 GHz} & 
\thead{$F_{\nu}$ 22 GHz} & 
\thead{\makecell{VLA \\ Config}}
}
\startdata
53027.65 &  & 1633 $\pm$ 54 &  &  & BC \\ \hline
53029.41 & 556 $\pm$ 55 & 1676 $\pm$ 42 &  &  & BC \\ \hline
53030.36 &  &  &  & 1582 $\pm$ 67 & BC \\ \hline
53041.36 & 938 $\pm$ 48 & 1893 $\pm$ 35 &  & 1856 $\pm$ 71 & BC \\ \hline
53069.4 &  & 4567 $\pm$ 38 &  & 5205 $\pm$ 96 & C \\ \hline
53081.15 & 2326 $\pm$ 106 & 5957 $\pm$ 54 &  & 5964  $\pm$ 167 & C \\ \hline
53083.25 &  &  & 8284 $\pm$ 153 &  & C \\ \hline
53088.21 & 3101 $\pm$ 78 & 7012 $\pm$ 58 & 9016 $\pm$ 219 & 6508 $\pm$ 164 & C \\ \hline
53101.07 & 3952 $\pm$ 102 &  & 9739 $\pm$ 232 &  & C \\ \hline
53107.14 & 3678 $\pm$ 77 & 7661 $\pm$ 59 & 10607 $\pm$ 234 & 7196 $\pm$ 151 & C \\ \hline
53113.08 & 3387 $\pm$ 78 & 8258 $\pm$ 59 & 10940  $\pm$ 256 & 7017 $\pm$ 172 & C \\ \hline
53127.14 & 4473 $\pm$ 82 & 7829 $\pm$ 62 & 10151 $\pm$ 219 & 7651 $\pm$ 134 & C \\ \hline
53137.18 & 4927 $\pm$ 80 & 9587 $\pm$ 61 & 9983 $\pm$ 247 & 7216 $\pm$ 157 & C \\ \hline
53154.16 &  & 8175 $\pm$ 91 & 7963 $\pm$ 212 & 6075 $\pm$ 130 & CD \\ \hline
53165.98 & 5574 $\pm$ 120 & 7942 $\pm$ 58 &  &  & CD \\ \hline
53166.04 &  &  & 6657 $\pm$ 203 & 4341 $\pm$ 91 & CD \\ \hline
53183.08 & 8097 $\pm$ 236 & 8889 $\pm$ 109 &  &  & D \\ \hline
53189.05 &  & 7613 $\pm$ 122 &  & 4267  $\pm$ 123 & D \\ \hline
53211.98 & 7808 $\pm$ 238 & 7575 $\pm$ 118 &  &  & D \\ \hline
53212.02 &  &  & 3818 $\pm$ 208 &  & D \\ \hline
53212.99 &  &  &  & 3682 $\pm$ 375 & D \\ \hline
53228.97 & 6505 $\pm$ 249 & 7156 $\pm$ 130 &  & 2654 $\pm$ 211 & D \\ \hline
53229.02 &  &  & 4488 $\pm$ 271 &  & D \\ \hline
53271.84 & 8873 $\pm$ 545 & 6084 $\pm$ 239 & 2300 $\pm$ 201 & 1722 $\pm$ 171 & A \\ \hline
53291.59 & 9423 $\pm$ 128 & 6487 $\pm$ 74 & 3850 $\pm$ 181 & 2816  $\pm$ 122 & A \\ \hline
53330.67 & 7796 $\pm$ 126 & 5091 $\pm$  120 & 2978 $\pm$ 198 & 1749 $\pm$ 118 & A \\ \hline
53393.2 &  & 5273 $\pm$ 254 &  &  & AB \\ \hline
53402.27 & 8267 $\pm$ 233 &  & 2270 $\pm$ 384 & 2058 $\pm$ 249 & AB \\ \hline
53511.47 & 8620 $\pm$ 271 &  &  &  & B \\ \hline
53511.9 &  & 5486 $\pm$ 455 & 2178 $\pm$ 231 & 1538 $\pm$ 166 & B \\ \hline
53603.94 & 7271 $\pm$ 367 & 3968 $\pm$ 187 & 2140 $\pm$ 671 & 1128 $\pm$ 658 & C \\ \hline
53689.6 & 10773 $\pm$ 365 & 5355 $\pm$ 231 & 4086 $\pm$ 239 & 2194 $\pm$ 238 & D \\ \hline
53801.41 & 5437 $\pm$ 139 & 3356 $\pm$ 183 & 1833 $\pm$ 251 & 1300 $\pm$ 135 & A \\ \hline
54248.1 & 3015 $\pm$ 116 & 1586 $\pm$ 15 &  & 737 $\pm$ 131 & A \\ \hline
54366.74 & 1751 $\pm$ 163 & 963 $\pm$ 77 &  &  & AB \\ \hline
54772.5 & 1226 $\pm$ 212 & 825 $\pm$ 100 &  &  & A \\ \hline
54919.0 & & $740 \pm 130$ & & & VLBI* \\ \hline
54926.22 & 1424 $\pm$ 78 & 721 $\pm$ 73 &  &  & B \\ \hline
55781.65 & 546 $\pm$ 60 &  &  &  & A \\ \hline
\enddata
\end{deluxetable}

\begin{deluxetable}{|c|l|l|l|l|l|l|l|}
\tablecaption{The following table reports the values plotted in Figures \ref{fig:brv}$-$\ref{fig:RhovsR} for $\epsilon_B=0.01$ and $\epsilon_e=0.1$, using the favored SED fit model BPL1. 
$\dot{M}$ is calculated assuming a wind velocity of 1000 \kms. 
The measurement taken 1931 days post explosion is the radius measurement from VLBI (described in Section \ref{sec:obs}). }
\label{tab:phys_vals}
\tablecolumns{8}
\tablewidth{\textwidth}
\tablehead{
\thead{ \makecell{Days Since \\ Explosion} } &
\thead{ $B$ (G)} &
\thead{ $U$ (erg) } &
\thead{$R_{\rm{FS}}$ (cm) } &
\thead{ $n_e$ (cm$^{-3}$)} &
\thead{ \makecell{$\dot{M}$ \\
($10^{-5}$\sm yr$^{-1}$)}} &
\thead{$v_{\rm{FS}}$ (cm s$^{-1}$) } &
\thead{$\rho_{\rm CSM}$ (g cm$^{-3}$)}}
\startdata
41.41 -- 42.36 & 0.561$\pm$0.035 & (1.4$\pm$0.2)$\times 10^{48}$ & (8.10$\pm$0.50)$\times 10^{15}$ & (2.5$\pm$0.6)$\times 10^5$ & 54$\pm$8 & (2.26$\pm$0.14)$\times 10^9$ & (4.1$\pm$0.9)$\times 10^{-19}$ \\
55.36 & 0.508$\pm$0.029 & (1.8$\pm$0.3)$\times 10^{48}$ & (9.40$\pm$0.50)$\times 10^{15}$ & (2.7$\pm$0.5)$\times 10^5$ & 79$\pm$12 & (1.97$\pm$0.11)$\times 10^9$ & (4.4$\pm$0.9)$\times 10^{-19}$ \\
82.40 & 0.533$\pm$0.030 & (5.0$\pm$0.8)$\times 10^{48}$ & (1.29$\pm$0.07)$\times 10^{16}$ & (3.5$\pm$0.7)$\times 10^5$ & 190$\pm$28 & (1.81$\pm$0.10)$\times 10^9$ & (5.8$\pm$1.2)$\times 10^{-19}$ \\
94.15 & 0.494$\pm$0.024 & (7.0$\pm$1.1)$\times 10^{48}$ & (1.51$\pm$0.07)$\times 10^{16}$ & (2.8$\pm$0.5)$\times 10^5$ & 220$\pm$29 & (1.85$\pm$0.09)$\times 10^9$ & (4.7$\pm$0.8)$\times 10^{-19}$ \\
101.21 & 0.454$\pm$0.021 & (8.6$\pm$1.3)$\times 10^{48}$ & (1.71$\pm$0.08)$\times 10^{16}$ & (2.2$\pm$0.4)$\times 10^5$ & 210$\pm$28 & (1.95$\pm$0.09)$\times 10^9$ & (3.6$\pm$0.6)$\times 10^{-19}$ \\
114.07 & 0.420$\pm$0.024 & (1.1$\pm$0.2)$\times 10^{49}$ & (1.96$\pm$0.10)$\times 10^{16}$ & (1.8$\pm$0.4)$\times 10^5$ & 230$\pm$33 & (1.99$\pm$0.10)$\times 10^9$ & (3.0$\pm$0.6)$\times 10^{-19}$ \\
120.14 & 0.444$\pm$0.020 & (1.0$\pm$0.2)$\times 10^{49}$ & (1.83$\pm$0.08)$\times 10^{16}$ & (2.5$\pm$0.4)$\times 10^5$ & 280$\pm$40 & (1.76$\pm$0.08)$\times 10^9$ & (4.3$\pm$0.7)$\times 10^{-19}$ \\
126.08 & 0.450$\pm$0.021 & (1.0$\pm$0.2)$\times 10^{49}$ & (1.81$\pm$0.08)$\times 10^{16}$ & (2.9$\pm$0.5)$\times 10^5$ & 320$\pm$40 & (1.70$\pm$0.10)$\times 10^9$ & (4.9$\pm$0.8)$\times 10^{-19}$ \\
140.14 & 0.417$\pm$0.019 & (1.1$\pm$0.2)$\times 10^{49}$ & (1.96$\pm$0.09)$\times 10^{16}$ & (2.6$\pm$0.4)$\times 10^5$ & 340$\pm$40 & (1.62$\pm$0.07)$\times 10^9$ & (4.4$\pm$0.7)$\times 10^{-19}$ \\
150.18 & 0.387$\pm$0.018 & (1.2$\pm$0.2)$\times 10^{49}$ & (2.15$\pm$0.10)$\times 10^{16}$ & (2.2$\pm$0.4)$\times 10^5$ & 340$\pm$40 & (1.66$\pm$0.08)$\times 10^9$ & (3.7$\pm$0.6)$\times 10^{-19}$ \\
167.16 & 0.406$\pm$0.029 & (9.6$\pm$1.8)$\times 10^{48}$ & (1.91$\pm$0.16)$\times 10^{16}$ & (3.8$\pm$1.1)$\times 10^5$ & 460$\pm$80 & (1.32$\pm$0.11)$\times 10^9$ & (6.3$\pm$1.9)$\times 10^{-19}$ \\
178.98 -- 179.04 & 0.315$\pm$0.016 & (1.1$\pm$0.2)$\times 10^{49}$ & (2.39$\pm$0.14)$\times 10^{16}$ & (1.7$\pm$0.4)$\times 10^5$ & 320$\pm$40 & (1.55$\pm$0.09)$\times 10^9$ & (2.8$\pm$0.6)$\times 10^{-19}$ \\
196.08 & 0.252$\pm$0.023 & (1.6$\pm$0.3)$\times 10^{49}$ & (3.14$\pm$0.30)$\times 10^{16}$ & (7.4$\pm$2.7)$\times 10^{4}$ & 240$\pm$50 & (1.85$\pm$0.18)$\times 10^9$ & (1.2$\pm$0.5)$\times 10^{-19}$ \\
202.05 & 0.330$\pm$0.040 & (1.0$\pm$0.3)$\times 10^{49}$ & (2.20$\pm$0.40)$\times 10^{16}$ & (2.7$\pm$1.6)$\times 10^{5}$ & 440$\pm$120 & (1.28$\pm$0.21)$\times 10^9$ & (4.4$\pm$2.6)$\times 10^{-19}$ \\
224.98 – 225.99 & 0.219$\pm$0.022 & (1.6$\pm$0.4)$\times 10^{49}$ & (3.40$\pm$0.40)$\times 10^{16}$ & (6.1$\pm$2.7)$\times 10^{4}$ & 240$\pm$50 & (1.77$\pm$0.22)$\times 10^9$ & (1.0$\pm$0.4)$\times 10^{-19}$ \\
241.97 -- 242.02 & 0.245$\pm$0.022 & (1.3$\pm$0.3)$\times 10^{49}$ & (2.96$\pm$0.32)$\times 10^{16}$ & (1.2$\pm$0.5)$\times 10^5$ & 350$\pm$70 & (1.42$\pm$0.15)$\times 10^9$ & (2.0$\pm$0.8)$\times 10^{-19}$ \\
284.84 & $\leq$0.255 & $\geq$8.4$\times 10^{48}$ & $\geq$2.49$\times 10^{16}$ & $\leq$2.5 $\times 10^5$ & $\leq$530 & $\geq$1.01$\times 10^9$ & $\leq$4.2$\times 10^{-19}$ \\
304.59 &$\leq$0.250 & $\geq$1.0$\times 10^{49}$ & $\geq$2.70$\times 10^{16}$ & $\leq$2.4$\times 10^5$ & $\leq$580 & $\geq$1.03$\times 10^9$ & $\leq$3.9$\times 10^{-19}$ \\
343.67 & $\leq$0.257 & $\geq$7.6$\times 10^{48}$ & $\geq$2.39$\times 10^{16}$ & $\leq$4.1$\times 10^5$ & $\leq$779 & $\geq$8.06$\times 10^8$ & $\leq$6.8$\times 10^{-19}$ \\
406.20 -- 416.27 & $\leq$0.256 & $\geq$8.0$\times 10^{48}$ & $\geq$2.44$\times 10^{16}$ & $\leq$5.7$\times 10^5$ & $\leq$1100 & $\geq$6.80$\times 10^8$ & $\leq$9.4$\times 10^{-19}$ \\
523.47 -- 523.90 &$\leq$0.257 & $\geq$7.7$\times 10^{48}$ & $\geq$2.40$\times 10^{16}$ & $\leq$9.4$\times 10^5$ & $\leq$1800 & $\geq$5.30$\times 10^8$ & $\leq$1.5$\times 10^{-18}$ \\
615.94 &$\leq$0.261 & $\geq$6.5$\times 10^{48}$ & $\geq$2.24$\times 10^{16}$ & $\leq$1.5$\times 10^6$ & $\leq$2600 & $\geq$4.22$\times 10^8$ & $\leq$2.5$\times 10^{-18}$ \\
701.60 &$\leq$0.250 & $\geq$1.0$\times 10^{49}$ & $\geq$2.69$\times 10^{16}$ & $\leq$1.3$\times 10^6$ & $\leq$3000 & $\geq$4.44$\times 10^8$ & $\leq$2.1$\times 10^{-18}$ \\
813.41 &$\leq$0.268 & $\geq$4.8$\times 10^{48}$ & $\geq$2.00$\times 10^{16}$ & $\leq$3.5$\times 10^6$ & $\leq$4700 & $\geq$2.84$\times 10^8$ & $\leq$5.9$\times 10^{-18}$ \\
1260.10 & $\leq$0.286 & $\geq$2.2$\times 10^{48}$ & $\geq$1.47$\times 10^{16}$ & $\leq$1.8$\times 10^7$ & $\leq$13000 & $\geq$1.35$\times 10^8$ & $\leq$2.9$\times 10^{-17}$ \\
1378.74 &$\leq$0.302 & $\geq$1.2$\times 10^{48}$ & $\geq$1.16$\times 10^{16}$ & $\leq$3.8$\times 10^7$ & $\leq$17000 & $\geq$9.80$\times 10^7$ & $\leq$6.3$\times 10^{-17}$ \\
1785.5 &$\leq$0.308 & $\geq$9.6$\times 10^{47}$ & $\geq$1.06$\times 10^{16}$ & $\leq$8.0$\times 10^7$ & $\leq$30000 & $\geq$6.90$\times 10^7$ & $\leq$1.3$\times 10^{-16}$ \\
1931.0 & -- & -- & \vlbaMeas & -- & -- &-- & -- \\
1938.22 &$\leq$0.309 & $\geq$9.2$\times 10^{47}$ & $\geq$1.04$\times 10^{16}$ & $\leq$9.8$\times 10^7$ & $\leq$36000 & $\geq$6.25$\times 10^7$ & $\leq$1.6$\times 10^{-16}$
\enddata
\end{deluxetable}

\begin{deluxetable}{|c|l|l|l|l|l|l|l|}
\tablecaption{The following table reports the values plotted in Figures \ref{fig:brv}$-$\ref{fig:RhovsR} for equipartition $\epsilon_B=\epsilon_e= \frac{1}{3}$, using the favored SED fit model BPL1. $\dot{M}$ assumes a wind velocity of 1000 $\kms$.
The measurement taken 1931 days post explosion is the radius measurement from VLBI (described in Section \ref{sec:obs}). }
\label{tab:phys_vals_equi}
\tablecolumns{8}
\tablewidth{\textwidth}
\tablehead{
\thead{ \makecell{Days Since \\ Explosion} } &
\thead{ $B$ (G)} &
\thead{ $U$ (erg) } &
\thead{$R_{\rm{FS}}$ (cm) } &
\thead{ $n_e$ (cm$^{-3}$)} &
\thead{\makecell{$\dot{M}$ \\ ($10^{-5}$ \sm\ yr$^{-1}$)}} &
\thead{$v_{\rm FS}$ (cm s$^{-1})$ } &
\thead{ $\rho_{\rm CSM}$ (g cm$^{-3})$}
}
\startdata
41.41 – 42.36 & 0.91$\pm$0.06 & (1.60$\pm$0.25)$\times10^{47}$ & (9.10$\pm$0.50)$\times10^{15}$ & (1.5$\pm$0.3)$\times10^{4}$ & 4.3$\pm$0.7 & (2.55$\pm$0.15)$\times10^{9}$ & (2.6$\pm$0.6)$\times10^{-20}$ \\
55.36 & 0.83$\pm$0.05 & (2.08$\pm$0.31)$\times10^{47}$ & (1.07$\pm$0.06)$\times10^{16}$ & (1.6$\pm$0.3)$\times10^{4}$ & 6.3$\pm$0.9 & (2.23$\pm$0.12)$\times10^{9}$ & (2.8$\pm$0.6)$\times10^{-20}$ \\
82.40 & 0.87$\pm$0.05 & (5.80$\pm$0.90)$\times10^{47}$ & (1.45$\pm$0.08)$\times10^{16}$ & (2.2$\pm$0.5)$\times10^{4}$ & 15$\pm$2.2 & (2.04$\pm$0.11)$\times10^{9}$ & (3.6$\pm$0.8)$\times10^{-20}$ \\
94.15 & 0.81$\pm$0.04 & (8.00$\pm$1.20)$\times10^{47}$ & (1.70$\pm$0.08)$\times10^{16}$ & (1.7$\pm$0.3)$\times10^{4}$ & 17$\pm$2.3 & (2.09$\pm$0.10)$\times10^{9}$ & (3.0$\pm$0.5)$\times10^{-20}$ \\
101.21 & 0.74$\pm$0.03 & (9.90$\pm$1.40)$\times10^{47}$ & (1.93$\pm$0.08)$\times10^{16}$ & (1.3$\pm$0.2)$\times10^{4}$ & 17$\pm$2.1 & (2.21$\pm$0.10)$\times10^{9}$ & (2.3$\pm$0.4)$\times10^{-20}$ \\
114.07 & 0.69$\pm$0.04 & (1.28$\pm$0.20)$\times10^{48}$ & (2.22$\pm$0.11)$\times10^{16}$ & (1.1$\pm$0.2)$\times10^{4}$ & 18$\pm$2.6 & (2.25$\pm$0.11)$\times10^{9}$ & (1.9$\pm$0.4)$\times10^{-20}$ \\
120.14 & 0.72$\pm$0.03 & (1.15$\pm$0.17)$\times10^{48}$ & (2.06$\pm$0.09)$\times10^{16}$ & (1.6$\pm$0.2)$\times10^{4}$ & 23$\pm$2.8 & (1.99$\pm$0.09)$\times10^{9}$ & (2.7$\pm$0.4)$\times10^{-20}$ \\
126.08 & 0.73$\pm$0.03 & (1.16$\pm$0.17)$\times10^{48}$ & (2.05$\pm$0.09)$\times10^{16}$ & (1.8$\pm$0.3)$\times10^{4}$ & 26$\pm$3.2 & (1.88$\pm$0.08)$\times10^{9}$ & (3.1$\pm$0.5)$\times10^{-20}$ \\
140.14 & 0.68$\pm$0.03 & (1.27$\pm$0.18)$\times10^{48}$ & (2.22$\pm$0.10)$\times10^{16}$ & (1.7$\pm$0.3)$\times10^{4}$ & 27$\pm$3.4 & (1.83$\pm$0.08)$\times10^{9}$ & (2.8$\pm$0.4)$\times10^{-20}$ \\
150.18 & 0.63$\pm$0.03 & (1.43$\pm$0.21)$\times10^{48}$ & (2.43$\pm$0.11)$\times10^{16}$ & (1.4$\pm$0.2)$\times10^{4}$ & 27$\pm$3.4 & (1.87$\pm$0.09)$\times10^{9}$ & (2.3$\pm$0.4)$\times10^{-20}$ \\
167.16 & 0.66$\pm$0.05 & (1.10$\pm$0.19)$\times10^{48}$ & (2.16$\pm$0.18)$\times10^{16}$ & (2.3$\pm$0.7)$\times10^{4}$ & 36$\pm$6 & (1.50$\pm$0.13)$\times10^{9}$ & (3.9$\pm$1.2)$\times10^{-20}$ \\
178.98 – 179.04 & 0.51$\pm$0.03 & (1.30$\pm$0.20)$\times10^{48}$ & (2.70$\pm$0.16)$\times10^{16}$ & (1.0$\pm$0.2)$\times10^{4}$ & 25.$\pm$3.4 & (1.75$\pm$0.10)$\times10^{9}$ & (1.7$\pm$0.4)$\times10^{-20}$ \\
196.08 & 0.41$\pm$0.04 & (1.89$\pm$0.34)$\times10^{48}$ & (3.55$\pm$0.34)$\times10^{16}$ & (4.6$\pm$1.7)e+03 & 19$\pm$4 & (2.10$\pm$0.20)$\times10^{9}$ & (7.7$\pm$2.8)$\times10^{-21}$ \\
202.05 & 0.54$\pm$0.07 & (1.15$\pm$0.30)$\times10^{48}$ & (2.50$\pm$0.40)$\times10^{16}$ & (1.7$\pm$1.0)$\times10^{4}$ & 35$\pm$10 & (1.44$\pm$0.23)$\times10^{9}$ & (2.8$\pm$1.6)$\times10^{-20}$ \\
224.98 – 225.99 & 0.35$\pm$0.03 & (1.90$\pm$0.40)$\times10^{48}$ & (3.90$\pm$0.50)$\times10^{16}$ & (3.8$\pm$1.7)e+03 & 19$\pm$4 & (2.00$\pm$0.24)$\times10^{9}$ & (6.4$\pm$2.8)$\times10^{-21}$ \\
241.97 – 242.02 & 0.39$\pm$0.03 & (1.49$\pm$0.29)$\times10^{48}$ & (3.30$\pm$0.40)$\times10^{16}$ & (7.5$\pm$2.9)e+03 & 28$\pm$5 & (1.60$\pm$0.17)$\times10^{9}$ & (1.2$\pm$0.5)$\times10^{-20}$ \\
284.84 & $\leq$0.41 & $\geq$9.72$\times10^{47}$ & $\geq$2.82$\times10^{16}$ & $\leq$1.5$\times10^{4}$ & $\leq$42 & $\geq$1.15$\times10^{9}$ & $\leq$2.6$\times10^{-20}$ \\
304.59 & $\leq$0.40 & $\geq$1.19$\times10^{48}$ & $\geq$3.06$\times10^{16}$ & $\leq$1.5$\times10^{4}$ & $\leq$46 & $\geq$1.16$\times10^{9}$ & $\leq$2.4$\times10^{-20}$ \\
343.67 & $\leq$0.42 & $\geq$8.72$\times10^{47}$ & $\geq$2.70$\times10^{16}$ & $\leq$2.6$\times10^{4}$ & $\leq$62 & $\geq$9.11$\times10^{8}$ & $\leq$4.2$\times10^{-20}$ \\
406.20 – 416.27 & $\leq$0.41 & $\geq$9.23$\times10^{47}$ & $\geq$2.76$\times10^{16}$ & $\leq$3.5$\times10^{4}$ & $\leq$90 & $\geq$7.69$\times10^{8}$ & $\leq$5.9$\times10^{-20}$ \\
523.47 – 523.90 & $\leq$0.41 & $\geq$8.82$\times10^{47}$ & $\geq$2.71$\times10^{16}$ & $\leq$5.9$\times10^{4}$ & $\leq$140 & $\geq$6.00$\times10^{8}$ & $\leq$9.8$\times10^{-20}$ \\
615.94 & $\leq$0.42 & $\geq$7.43$\times10^{47}$ & $\geq$2.54$\times10^{16}$ & $\leq$9.5$\times10^{4}$ & $\leq$200 & $\geq$4.77$\times10^{8}$ & $\leq$1.5$\times10^{-19}$ \\
701.60 & $\leq$0.41 & $\geq$1.17$\times10^{48}$ & $\geq$3.04$\times10^{16}$ & $\leq$8.0$\times10^{4}$ & $\leq$250 & $\geq$5.01$\times10^{8}$ & $\leq$1.3$\times10^{-19}$ \\
813.41 & $\leq$0.43 & $\geq$5.53$\times10^{47}$ & $\geq$2.26$\times10^{16}$ & $\leq$2.2$\times10^{5}$ & $\leq$380 & $\geq$3.22$\times10^{8}$ & $\leq$3.6$\times10^{-19}$ \\
1260.10 & $\leq$0.46 & $\geq$2.56$\times10^{47}$ & $\geq$1.67$\times10^{16}$ & $\leq$1.1$\times10^{6}$ & $\leq$1000 & $\geq$1.53$\times10^{8}$ & $\leq$1.8$\times10^{-18}$ \\
1378.74 & $\leq$0.49 & $\geq$1.40$\times10^{47}$ & $\geq$1.31$\times10^{16}$ & $\leq$2.3$\times10^{6}$ & $\leq$1400 & $\geq$1.11$\times10^{8}$ & $\leq$3.9$\times10^{-18}$ \\
1785.5 & $\leq$0.50 & $\geq$1.11$\times10^{47}$ & $\geq$1.20$\times10^{16}$ & $\leq$5.0$\times10^{6}$ & $\leq$2400 & $\geq$7.80$\times10^{7}$ & $\leq$8.3$\times10^{-18}$ \\
1931.0 & -- & -- & \vlbaMeas & -- & -- &-- & -- \\
1938.22 & $\leq$0.50 & $\geq$1.06$\times10^{47}$ & $\geq$1.18$\times10^{16}$ & $\leq$6.1$\times10^{6}$ & $\leq$2900 & $\geq$7.07$\times10^{7}$ & $\leq$1.0$\times10^{-17}$
\enddata
\end{deluxetable}